\documentclass[traditabstract]{aa} 

\usepackage[colorlinks=true, linkcolor=blue, citecolor=blue, urlcolor=blue]{hyperref}
\usepackage{graphicx}
\usepackage{txfonts}
\usepackage{natbib}

\begin{document}

   \title{TESS unveils the phase curve of WASP-33b}
   \subtitle{Characterization of the planetary atmosphere and the pulsations from the star}
 \author{C. von Essen\inst{1,2}
          \and
          M. Mallonn\inst{3}
          \and
          C. C. Borre\inst{1}
          \and
          V. Antoci\inst{4,1}
          \and
          K.G. Stassun\inst{5}
          \and
          S. Khalafinejad\inst{6}
          \and
          G. Tautvai{\v s}ien{\.e}\inst{2}
          }

   \institute{Stellar Astrophysics Centre, Department of Physics and Astronomy, Aarhus University, Ny Munkegade 120, DK-8000 Aarhus C, Denmark\\
         \email{cessen@phys.au.dk}
         \and
             Astronomical Observatory, Institute of Theoretical Physics and Astronomy, Vilnius University, Sauletekio av. 3, 10257, Vilnius, Lithuania
         \and
             Leibniz-Institut f\"{u}r Astrophysik Potsdam (AIP), An der Sternwarte 16, D-14482 Potsdam, Germany
        \and
            DTU Space, National Space Institute, Technical University of Denmark, Elektrovej 328, DK-2800 Kgs. Lyngby, Denmark
        \and 
            Vanderbilt University, Department of Physics \& Astronomy, 6301 Stevenson Center Ln., Nashville, TN 37235, USA
        \and 
            Landessternwarte, Zentrum f\"{u}r Astronomie der Universit\"{a}t Heidelberg, K\"{o}nigstuhl 12, 69117 Heidelberg, Germany
             }
   \date{Received 09.03.2020; accepted 22.04.2020}
   
\abstract{We present the detection and characterization of the
  full-orbit phase curve and secondary eclipse of the ultra-hot
  Jupiter WASP-33b at optical wavelengths, along with the pulsation
  spectrum of the host star. We analyzed data collected by the
  Transiting Exoplanet Survey Satellite (TESS) in sector 18. WASP-33b
  belongs to a very short list of highly irradiated exoplanets that
  were discovered from the ground and were later visited by TESS. The
  host star of WASP-33b is of $\delta$\,Scuti-type and shows nonradial
  pulsations in the millimagnitude regime, with periods comparable to
  the period of the primary transit. These completely deform the
  photometric light curve, which hinders our interpretations. By
  carrying out a detailed determination of the pulsation spectrum of
  the host star, we find 29 pulsation frequencies with a
  signal-to-noise ratio higher than 4. After cleaning the light curve
  from the stellar pulsations, we confidently report a secondary
  eclipse depth of \mbox{305.8 $\pm$ 35.5} parts-per-million (ppm),
  along with an amplitude of the phase curve of \mbox{100.4 $\pm$ 13.1
    ppm} and a corresponding westward offset between the region of
  maximum brightness and the substellar point of \mbox{28.7 $\pm$ 7.1}
  degrees, making WASP-33b one of the few planets with such an offset
  found so far. Our derived Bond albedo, \mbox{A$_B$ = 0.369 $\pm$
    0.050}, and heat recirculation efficiency, \mbox{$\epsilon$ =
    0.189 $\pm$ 0.014}, confirm again that he behavior of WASP-33b is
  similar to that of other hot Jupiters, despite the high irradiation
  received from its host star. By connecting the amplitude of the
  phase curve to the primary transit and depths of the secondary
  eclipse, we determine that the day- and nightside brightness
  temperatures of WASP-33b are \mbox{3014 $\pm$ 60 K} and \mbox{1605
    $\pm$ 45 K}, respectively. From the detection of photometric
  variations due to gravitational interactions, we estimate a planet
  mass of $M_{P}=2.81 \pm 0.53$~M$_\mathrm{J}$. Based on analyzing the
  stellar pulsations in the frame of the planetary orbit, we find no
  signals of star-planet interactions.}

\keywords{stars: planetary systems -- stars: individual: WASP-33 -- methods: observational}

\maketitle

\section{Introduction}

The Transiting Exoplanet Survey Satellite \citep[TESS,][]{Ricker2015}
has been scanning the southern and northern ecliptic hemispheres since
August 2018 in the search for planets around bright stars. To date
(April 2020), TESS has detected the dimming of light during transit of
$\sim$1800 TESS objects of interest (TOIs), about 45 of which have
been confirmed as
exoplanets\footnote{\url{https://tess.mit.edu/publications/}}. Several
TESS discoveries include the first Earth-sized planet
\citep{Dragomir2019}, an eccentric massive Jupiter orbiting a subgiant
star every 9.5 days \citep{Rodriguez2019}, and the first multiplanet
systems \citep{Kostov2019,Gunther2019,Vanderburg2019}.

In addition to the detection and characterization of new systems, TESS
has also been contributing with the in-depth study of systems
previously detected from the ground. Some TESS contributions are the
detection of a decrease in the orbital period of \mbox{WASP-4b}
\citep{Bouma2019}, and of particular interest to this work, the
characterization of the phase curve and secondary eclipse depth of
\mbox{WASP-18b} \citep{Shporer2019}, \mbox{WASP-19b} \citep{Wong2020},
\mbox{WASP-121b} \citep{Bourrier2019,Daylan2019}, \mbox{WASP-100b}
\citep{Jansen2020}, and \mbox{KELT-9b} \citep{Wong2019b}. The
precision in the photometry of \mbox{WASP-18b} allowed
\cite{Shporer2019} to unveil sinusoidal modulations across the orbital
phase that were shaped by the atmospheric characteristics of the
planet and by the gravitational interactions between the planet and
host star. Data for WASP-19b revealed a strong atmospheric brightness
modulation signal and no significant offset detected between the
substellar point and the region of maximum brightness on the dayside
of the planet, in full agreement with data for \mbox{WASP-121b}.

All these planets with full-orbit phase curves measured by TESS belong
to the group of ultra-hot Jupiters. These planets receive such an
extreme amount of stellar insolation that they exhibit dayside
temperatures exceeding $\sim$2200~K. Similar to hot Jupiters of more
moderate temperatures, they are expected to be tidally locked because
of their very close proximity to the host star. Another ultra-hot
Jupiter observed by TESS is WASP-33b \citep{CollierCameron2010}. With
a dayside temperature close to 3200~K \citep{Zhang2018}, it belongs to
the very top of a temperature ranking of highly irradiated super-hot
exoplanets. The planet orbits a $\delta$ Scuti star of spectral type A
that oscillates with pulsations commensurable to the transit duration
and with amplitudes well within the millimagnitude regime
\citep{Smith2011,Herrero2011,vonEssen2014}. Relevant stellar and
planetary parameters can be found in Table~\ref{tab:third_light} and
Table~\ref{tab:transit_parameters}. So far, the planet has been
thoroughly investigated. Among others, observational data have
revealed several secondary eclipse depths at different wavelengths
\citep{Smith2011,Deming2012,deMooij2013,Haynes2015,vonEssen2015}, a
detailed characterization of the pulsation spectrum of the host star
with the goal of determining planetary parameters from
pulsation-cleaned light curves \citep{Herrero2011,vonEssen2014}, and
the characterization of its atmospheric composition where aluminium
oxide was unveiled for the first time \citep{vonEssen2019b}. In
addition to this detection, \cite{Yan2019} characterized its
transmission spectrum around the individual lines of Ca II H\&K in
high resolution, finding the spectrum mostly ionized in its upper
atmosphere, while \cite{Nugroho2017} found molecular TiO in the
dayside spectra. Space-based Spitzer observations of the WASP-33 phase
curves in the near-infrared (NIR) allowed \citep{Zhang2018} to
estimate the planetary brightness temperature, albedo, and heat
recirculation efficiency. The authors found that \mbox{WASP-33b}
shares similarities with hot Jupiters, despite its unusually high
irradiation level.

Phase-curve observations at TESS optical wavelengths allow measuring
the combined reflected and thermally emitted planetary light as a
function of longitude. Recent reviews of exoplanet phase curves were
provided by \cite{Shporer2017} and \cite{Parmentier2018}. For an
ultra-hot Jupiter such as WASP-33b, we still expect the thermal light
component to dominate, which is informative for the efficiency of the
heat distribution from the insolated dayside to the nightside. This
energy transport is expected to be less efficient for ultra-hot
Jupiters than for hot Jupiters of more moderate temperature because
radiative energy loss is stronger and the atmosphere is partially
ionized, which prevents strong advective energy transport by a
magnetic drag \citep[e.g.,][]{Lothringer2018,Arcangeli2019}. The
optical phase curve of WASP-33b, presented in this work, sheds light
light on the energy recirculation by providing the related observing
parameters of day-night temperature contrast and phase-curve offset
for an extremely highly irradiated planet. Moreover, a comparison of
the optical phase-curve results of this work and the NIR phase-curve
results for the same planet as described by \cite{Zhang2018} will
provide information about their wavelength dependence, and therefore
about how similar the planet atmosphere is to that of a
blackbody.

We describe in Sect.~\ref{sec:obs} the observations we used to
characterize the secondary eclipse and the phase curve of WASP-33b in
detail. In Sect.~\ref{sec:results} we present our analysis of the
third-light contamination (Sect.~\ref{sec:third_light}), our update on
the transit parameters (Sect.~\ref{sec:PT}), and our strategy of
cleaning the light curves from transits to use them to determine the
pulsation spectrum of the host star (Sect.~\ref{sec:PSpec}). We
introduce our models for interpreting the secondary eclipse and phase
curve of WASP-33b in Sect.~\ref{sec:SE_PC}. In
Section~\ref{sec:discussion} we discuss the effect of the pulsations
on the derived atmospheric parameters in detail
(Sect.~\ref{sec:Puls_Impact}), derive relevant physical parameters in
Sections~\ref{sec:params_W33_derived} and
\ref{sec:params_W33_derived_2}, place WASP-33b and our results in
context in Section~\ref{sec:context}, and update the stellar and
planetary parameters we presented in
Section~\ref{sec:updated_params}. We conclude in
Sect.~\ref{sec:conclusion}.

\section{Observations}
\label{sec:obs}

WASP-33 (TIC identifier 129979528) was observed by TESS in sector 18,
more specifically, it was observed between November 3$^{}$ and
26$^{}$, 2019, during cycle 2. Camera 1 was used. The data have a
cadence of 120 seconds and were analyzed and detrended by the Science
Processing Operations Center (SPOC) pipeline, based on the NASA Kepler
mission pipeline \citep{Jenkins2016,Jenkins2017}. Time stamps are
given in Barycentric Julian Dates (BJD$_\mathrm{TDB}$), and are
therefore not converted into another time-reference frame.

\begin{figure*}[ht!]
    \centering
    \includegraphics[width=\textwidth]{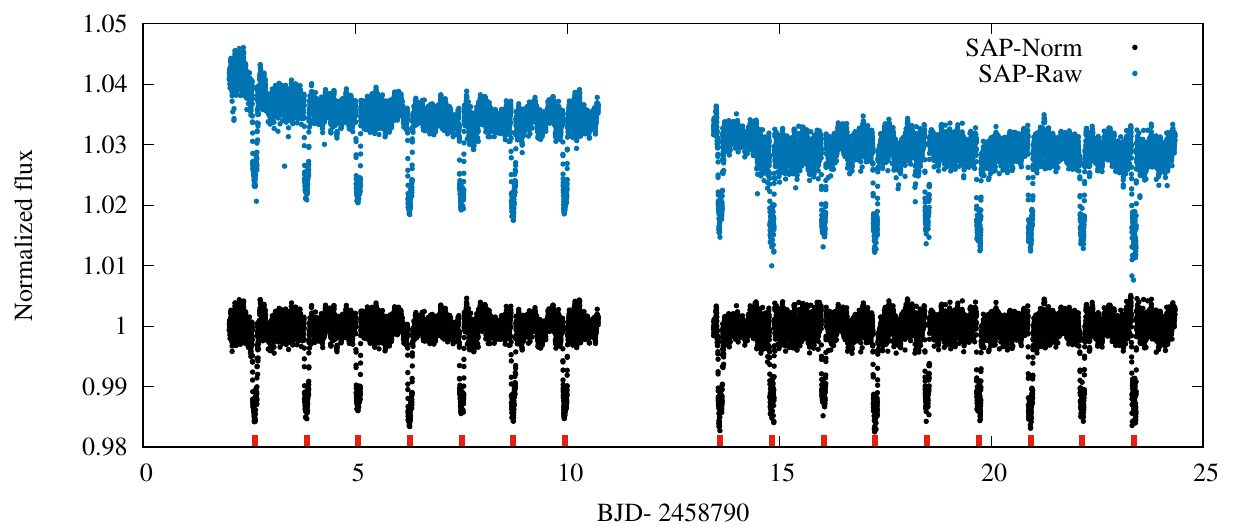}
    \caption{\label{fig:W33LC} SAP normalized flux of WASP-33 observed
      by TESS shown as black circles, along with SAP raw data in
      blue. The pulsations of the star deform the continuum level. The
      16 transits are indicated at the bottom with red lines. The gap
      in the middle is caused by data downlink dead time.}
\end{figure*}

The light curve of \mbox{WASP-33} is shown in
Fig.~\ref{fig:W33LC}. The total time on target is of about 23 days,
during which 16 primary transits were observed. The first $\sim$800
data points were not considered in our analysis because they show some
noise structure that is probably extrinsic to the star. An initial
analysis was made on the simple aperture photometry (SAP) and
presearch data conditioning (PDC) light curves
\citep{Smith2012,Stumpe2014}, after which we decided to work on the
SAP data using our own normalization strategy (see
Section~\ref{sec:noise} for a motivation of our choice). To prepare
the data for analysis, we first removed all flag points, both in time
and in flux. Then, we binned the points each 8 hours to minimize the
normalization effect on the stellar pulsations, after which we
interpolated a spline function. We used this to normalize the data and
to remove outliers that were five times the standard deviation
away. We removed about 100, and we analyzed in this work 14000 points
in total.

\section{Analysis and model considerations}
\label{sec:results}

\subsection{Third-light contamination}
\label{sec:third_light}

\begin{figure}[ht!]
    \centering
    \includegraphics[width=.5\textwidth]{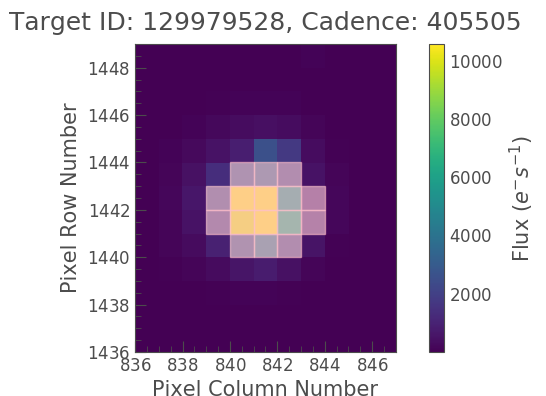}
    \includegraphics[width=.35\textwidth]{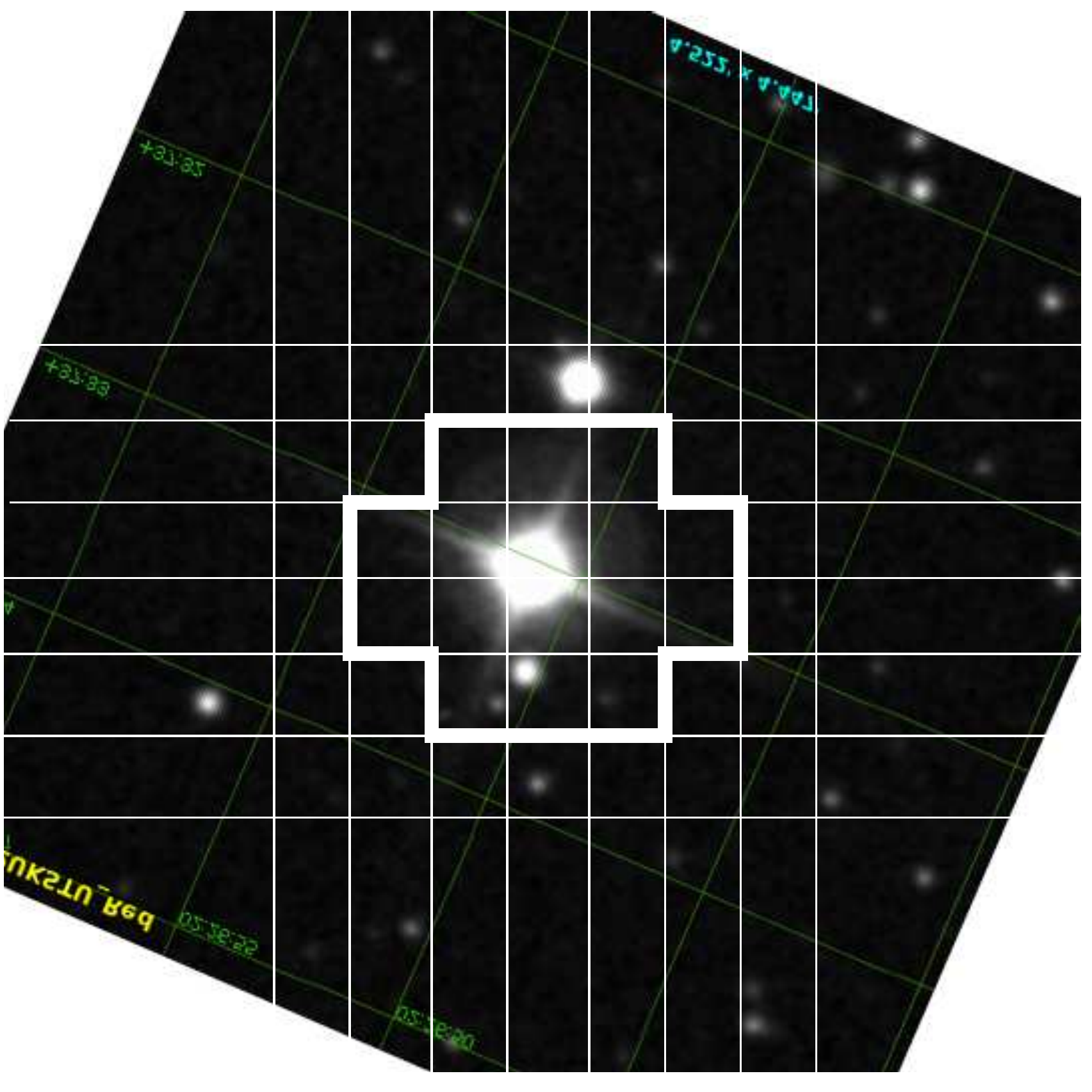}
    \caption{\label{fig:W33_aperture} {\it Top:} TPF of WASP-33
      showing the chosen aperture mask. No stars can be visually
      resolved. {\it Bottom:} Field of view of about 4$\times$4~arcmin
      centered on WASP-33. The mask and pixels are schematized with
      white thick and thin lines, respectively. WASP-33C, the bright
      star south of WASP-33, is not included in the aperture. The
      field of view has been rotated to be aligned with the ecliptic
      system, and is oriented to coincide with the images given in the
      TESS summary report.}
\end{figure}

When photometric time series are analyzed including exoplanetary
primary transits, special care has to be taken. In certain cases,
light of another star than the planetary host is included inside the
chosen photometric aperture, which dilutes the depth of the primary
transits \citep[see, e.g.,][for a large systematic search of
  companions of nearby
  exoplanets]{Piskorz2015,Mugrauer2019,Belokurov2020}. The TESS
cameras have a pixel size of 21$\times$21~arcseconds. Under these
circumstances, when a light curve is constructed by coadding the light
of several pixels (see Fig.~\ref{fig:W33_aperture}, {\it top}), it is
very likely that the aperture will include light from other stars than
that of the host (Figure~\ref{fig:W33_aperture}, {\it bottom}).

The first identified companion of WASP-33b, WASP-33B, lies at an
angular separation of $\sim$2~arcseconds and is therefore included
inside the TESS aperture. The substellar object was first reported by
\cite{Moya2011} and then confirmed by
\citet{Adams2013,Wollert2015,Ngo2016}. Using years of follow-up
observations, \cite{Ngo2016} carried out a combined analysis and
determined that WASP-33 is a binary system candidate. Today, WASP-33
has been identified as a hierarchical triple star system
\citep{Mugrauer2019} with a second companion, WASP-33C, that
is$\sim$49~arcseconds away from the planet-host star. Owing to its
orientation (southeast of WASP-33), WASP-33C is in principle not
included inside the TESS aperture. However, because of the large point
spread function of TESS, some light of WASP-33C might be
included. Unfortunately, there is no way to quantify this except for
comparing the derived transit depth to literature values and/or those
expected from atmospheric models. A third star, located 23~arcseconds
northwest of WAP-33 \citep[\mbox{Gaia DR2
    328636024020571008},][\mbox{G = 14.6173 $\pm$
    0.0005}]{GaiaCollaboration2018}, is included in the aperture. Even
though the Science Processing Operations Center (SPOC) pipeline
provides an estimate of the stellar crowding contamination, we thought
it prudent to compare this with the estimate determined from our own
analysis. Before the transit fitting, we therefore computed the
third-light contribution of the WASP-33 companions within the TESS
transmission response, that is, the contribution of WASP-33B and
\mbox{Gaia DR2 328636024020571008}, and compared this to the value
reported in the header of the WASP-33 fits file, specifically under
the CROWDSAP keyword.

Because of the nature of the system, we can assume that WASP-33 and
WASP-33B are at the same distance. We therefore reproduced their
emission with PHOENIX synthetic spectra \citep{phoenix} without the
need of scaling the fluxes further to account for
distances. Specifically, we used PHOENIX spectra with basic stellar
parameters ($T_{\rm eff}$, [Fe/H]. log($g$)) that match those of
WASP-33 \citep{CollierCameron2010} and of the close-in companion
\citep{Ngo2016}, as summarized in Table~\ref{tab:third_light}.

After convolving PHOENIX intensities with the TESS transmission
response, we integrated the remaining fluxes and computed their
ratio. In this way, we obtain a third-light contribution of WASP-33B
of \mbox{$F_{\rm W33B}$/F$_{\rm W33}$ = 0.018}.

The case of \mbox{Gaia DR2 328636024020571008} is slightly different
because it is not bound to WASP-33 by gravity, therefore we cannot
assume equal distances. \cite{GaiaCollaboration2018} estimated its
temperature to be \mbox{$\sim$5075 K}, therefore we represented its
emission using PHOENIX synthetic spectra for a main-sequence star of
\mbox{$T_{\rm eff} = 5000$~K}. To compensate for the difference in
distance, we computed the Gaia magnitude difference between WASP-33
\mbox{(G = 8.0700 $\pm$ 0.0004)} and this star, and we scaled PHOENIX
flux ratios of WASP-33 and \mbox{Gaia DR2 328636024020571008}
integrated within the Gaia transmission response to meet the magnitude
difference. Then, we used this factor to scale the spectra inside the
TESS transmission response down. In this way, the third-light
contribution of \mbox{Gaia DR2 328636024020571008} was found to be
\mbox{$F_{\rm Gaia}$/F$_{\rm W33}$ = 0.006}. The total third-light
contribution we used in our model is the addition of these two, and
equal to \mbox{$\Delta$F = 0.024}. In comparison, the value reported
at CROWDSAP is 0.9789, equivalently 1 - 0.9789 = 0.0211, to be
compared to our $\Delta$F. Because these two values differ by only
$\sim$10\%, we find the two values compatible, and we use our derived
value to correct for third light throughout.

\begin{table*}[ht!]
    \caption{\label{tab:third_light} Effective temperature, metallicity, and surface gravity for WASP-33 and the stars included in the TESS aperture.}
    \centering
    \begin{tabular}{l c c c}
    \hline\hline    
    Parameter             &    WASP-33                 &   WASP-33B              &  \mbox{Gaia DR2 328636024020571008}   \\
                          & \citep{CollierCameron2010} &  \citep{Ngo2016}        &  \cite{GaiaCollaboration2018}  \\
    \hline
    $T_{\rm eff}$ (K)     &    7430 $\pm$ 100          &    3050 $\pm$ 250       & 5074.75          \\ 
    $\mathrm{[Fe/H]}$     & 0.1 $\pm$ 0.2              & 0 (adopted)            &  0 (adopted)           \\
    log($g$)              &    4.3 $\pm$ 0.2           & 5 \citep{Angelov1996}   & 4.5 \citep{Angelov1996} \\
    \hline
    \end{tabular}
\end{table*}

\subsection{Limb-darkening coefficients}

We adopted a quadratic limb-darkening law,
\begin{equation}
\frac{I(\mu)}{I(1)} = 1 - u_1(1-\mu) - u_2(1-\mu)^2,
\label{eq_qua}
\end{equation}

\noindent with corresponding linear ($u_1$) and quadratic ($u_2$)
limb-darkening coefficients (LDCs). In the equation, $I(1)$ is the
specific intensity at the center of the stellar disk and $\mu =
\cos(\gamma)$, where $\gamma$ is the angle between the line of sight
and the emergent intensity. To compute our custom limb-darkening
coefficients that meet the TESS transmission response, we used
angle-dependent specific intensity spectra from PHOENIX
\citep{phoenix} with main stellar parameters corresponding to the
effective temperature, \mbox{$T_{\rm eff} = 7400$}~K, surface gravity,
\mbox{log($g$) = 4.5}, and metallicity, \mbox{[Fe/H] = 0.00}. This
matches the values of WASP-33 reported in Table~\ref{tab:third_light}
within the uncertainties. Following \cite{vonessen2017} and
\cite{claret2011}, we neglected the data points between \mbox{$\mu$ =
  0} and \mbox{$\mu$ = 0.07} because the intensity drop given by
PHOENIX models is too steep and might be unrealistic. After
integrating the PHOENIX angle-dependent spectra convolved by the TESS
response, we fit the derived intensities normalized by their maximum
values with Equation~\ref{eq_qua} using a Markov chain Monte Calro
(MCMC) approach. The derived limb-darkening coefficients for WASP-33
are \mbox{$u_1$ = 0.246(6)} and \mbox{$u_2$ = 0.252(6)}. Errors for
the coefficients were derived from the posterior distributions of the
MCMC chains after visually inspecting them for convergency. In order
to assess the quality of our procedure, we fit the LDCs to TESS
primary transit light curves. From their posterior distributions we
obtained consistent results with their PHOENIX counterparts. Our
derived LDCs agree well with those from \cite{Claret2017} (\mbox{$u_1$
  = 0.2446} and \mbox{$u_2$ = 0.2449}). A word of caution has to be
given here. Even though the precision of our fit lies in the fourth
decimal, LDCs drag uncharacterized errors from the lack of precision
of, for instance, stellar intensities. The precision of the fit
therefore probably reflects the real precision at which we know any
limb-darkening coefficient. Based on this assumption, we consider
$u_1$ and $u_2$ as fixed in order to reduce the computational cost in
this work.

\subsection{Pulsation spectrum of the host star}
\label{sec:PSpec}

\begin{figure*}[ht!]
    \centering
    \includegraphics[width=\textwidth]{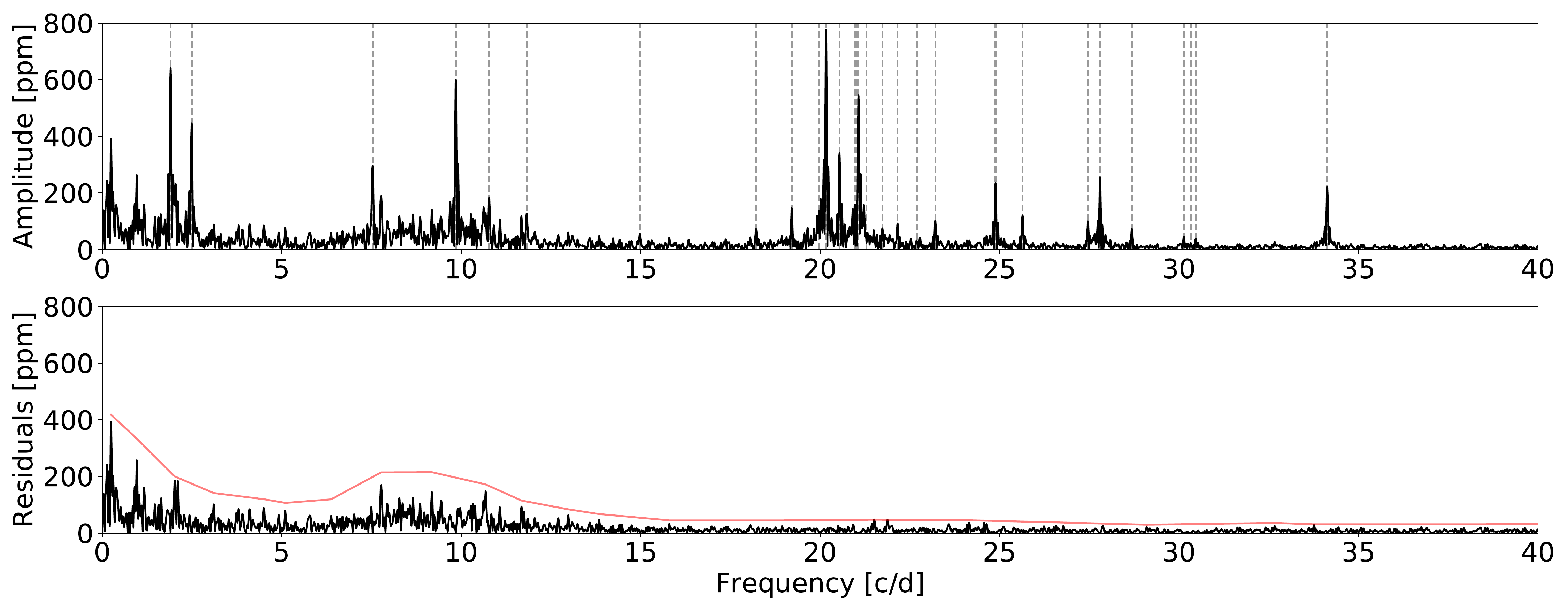}
    \caption{Top panel: Power spectrum of WASP-33. Pulsations are
      marked with dashed gray lines. Bottom panel: Residuals after
      frequency extraction. The red line makes the S/N limit of 4.}
    \label{fig:powerspectrum}
\end{figure*}

The pulsation frequencies of the host star were determined using the
code Period04 \citep{Lenz2005}, after the primary transits were fit
and removed. The complete procedure was carried out over the SAP and
PDC light curves, from which we obtained consistent results. The
software uses a fast Fourier transform to calculate the power spectrum
and simultaneous least-squares fitting to derive the pulsation
frequencies, their amplitudes, and phases. The pulsation frequencies
were extracted one by one, starting with the one having the largest
amplitude. We considered a peak to be statistically significant only
if it was resolved, and if its corresponding signal-to-noise ratio
(S/N) was higher than or equal to 4 \citep[see,
  e.g.,][]{Breger1993,vonEssen2014}. The S/N was calculated using the
default setting, that is, a window of two cycles/day (cd$^{-1}$)
around each peak. We estimated the uncertainties using the MCMC tool
of Period04. These are produced as described in \cite{Breger1999a},
are given at a 1$\sigma$ level and are derived from 1000 MCMC
iterations. The amplitude uncertainties depend only on the residuals
and the number of data points in the time series, which explains why
all the values are the same.

The power spectrum of WASP-33 is shown in
Fig.~\ref{fig:powerspectrum}, and the 29 extracted pulsation
frequencies, with their associated amplitudes and phases, are given in
Table~\ref{tab:pulsations}. For completeness, we provide a comparison
to the peaks found by \cite{vonEssen2014}.

Two frequencies identified by \cite{vonEssen2014} were not reproduced
in this analysis: \mbox{8.308 cd$^{-1}$} (Puls$_7$) and \mbox{10.825
  cd$^{-1}$} (Puls$_8$). Both are located in a frequency range whose
noise level in the TESS data is slightly higher, which is likely to be
due to additional unresolved pulsation modes originating from the
star. As a consequence, it is not clear whether these peaks now have a
lower amplitude than before and are therefore buried in the noise, or
whether the peaks identified before may have been the result of
aliases. We note that $\delta$~Sct-type pulsations are known to show
(sometimes strong) amplitude variability over time \citep[see,
  e.g.,][]{bowman2018}. In addition, the TESS bandpass is redder than
the filters used in \cite{vonEssen2014}, which implies that the
pulsation amplitudes are expected to be lower as well, also depending
on the exact geometry of the mode.  WASP-33 displays p-mode
oscillations at high frequencies, which is characteristic for
$\delta$\,Scuti stars \citep[see, e.g.,][]{Aerts2010,
  Antoci2019}. Owing to the 23 days of continuous monitoring provided
by TESS data, we also detected statistically significant peaks at
lower frequencies (F2 and F5 in Table~\ref{tab:pulsations}). If these
are independent pulsation modes, they would correspond to g-mode
pulsations typical for $\gamma$\,Doradus stars \citep[see,
  e.g.,][]{Aerts2010, gangli2020}. Showing both g- and p-mode
pulsations, WASP-33 would then fall into the
$\gamma$\,Doradus-$\delta$\,Scuti hybrid classification
\citep{Grigahcene2010,Balona2011,Uytterhoeven2011}. However, with only
two peaks detected at low frequencies, we prefer to wait for more data
before classifying WASP-33 as such. \cite{vonEssen2014} did not detect
signals with frequencies lower than $\sim$7 cd$^{-1}$ . In addition to
the obvious gaps produced by the day-night cycle, the data were
normalized on a nightly basis, which removed the long
trends. Moreover, as a result of poor weather conditions, the
observations could not be produced in consecutive nights, and on
average had a duration of $\sim$5 hours. A longer time series than the
one provided by TESS might resolve the pulsations at lower frequencies
and thereby determine whether the star is purely a $\delta$\,Scuti or
a hybrid star.

\begin{table*}[ht!]
        \centering
        \caption{\label{tab:pulsations} Pulsation frequencies of
          WASP-33 derived from TESS photometry. From left to right, we
          present the frequency number, F\#, arranged in decreasing
          amplitude, the frequency, in cd$^{-1}$, the amplitude, in
          ppm, the phase, in units of 2$\pi$, and the frequency, in
          cP$^{-1}$. In all cases, errors are given at 1$\sigma$
          level. The last column shows the frequencies obtained in
          \cite{vonEssen2014}, which agree with those found here.}
        \begin{tabular}{l l l l c c}
                \hline
                F\# & Frequency     & Amplitude                 & Phase     &  Frequency  &  Frequency \\
                &   (cd$^{-1}$)       & (ppm)                         & (2$\pi$)  &   (cP$^{-1}$)   &  \citep[][cd$^{-1}$]{vonEssen2014} \\
                \hline
F1 &    20.16263        $\pm$   0.00032 &       772     $\pm$   10      &       0.4904  $\pm$   0.0020 & 24.5957 $\pm$ 0.0004 &        20.16214 $\pm$ 0.00063 (Puls$_1$) \\
F2 &    1.89739 $\pm$   0.00038 &       648     $\pm$   10      &       0.0084  $\pm$   0.0024 & 2.3145 $\pm$ 0.0004 & -       \\
F3 &    9.84567 $\pm$   0.00041 &       604     $\pm$   10      &       0.4927  $\pm$   0.0026 & 12.0104 $\pm$ 0.0005 &        9.84361 $\pm$ 0.00066 (Puls$_3$)        \\
F4 &    21.06527        $\pm$   0.00043 &       564     $\pm$   10      &       0.3151  $\pm$   0.0028 & 25.6968 $\pm$ 0.0005 &        21.06057 $\pm$ 0.00058 (Puls$_2$)       \\
F5 &    2.48691 $\pm$   0.00052 &       468     $\pm$   10      &       0.6578  $\pm$   0.0034 & 3.0337 $\pm$ 0.0006 & -       \\
F6 &    20.53605        $\pm$   0.00073 &       334     $\pm$   10      &       0.2175  $\pm$   0.0047 & 25.0512 $\pm$ 0.0008 &        20.53534 $\pm$ 0.00057 (Puls$_5$)       \\
F7 &    7.52946 $\pm$   0.00083 &       296     $\pm$   10      &       0.5895  $\pm$   0.0053 & 9.1849 $\pm$ 0.0010 & -       \\
F8 &    27.79525        $\pm$   0.00096 &       256     $\pm$   10      &       0.9856  $\pm$   0.0062 & 33.9065 $\pm$ 0.0011 &        -       \\
F9 &    24.8835 $\pm$   0.0010  &       243     $\pm$   10      &       0.9808  $\pm$   0.0065 & 30.3545 $\pm$ 0.0012 &        24.88351 $\pm$ 0.00056 (Puls$_4$)       \\
F10 &   34.1254 $\pm$   0.0011  &       220     $\pm$   10      &       0.9392  $\pm$   0.0072 & 41.6284 $\pm$ 0.0013 &        34.12521 $\pm$ 0.00054 (Puls$_6$)       \\
F11 &   20.9668 $\pm$   0.0012  &       198     $\pm$   10      &       0.9824  $\pm$   0.0080 & 25.5767 $\pm$ 0.0014 &        -       \\
F12 &   10.7773 $\pm$   0.0013  &       184     $\pm$   10      &       0.0815  $\pm$   0.0086 & 13.1468 $\pm$ 0.0015 &        -       \\
F13 &   11.8238 $\pm$   0.0019  &       130     $\pm$   10      &       0.385   $\pm$   0.012 & 14.4234 $\pm$ 0.0023 & -      \\
F14 &   25.6394 $\pm$   0.0021  &       118     $\pm$   10      &       0.854   $\pm$   0.013 & 31.2766 $\pm$ 0.0025 & -      \\
F15 &   19.2058 $\pm$   0.0021  &       116     $\pm$   10      &       0.986   $\pm$   0.014 & 23.4285 $\pm$ 0.0025 & -      \\
F16 &   23.2070 $\pm$   0.0023  &       107     $\pm$   10      &       0.075   $\pm$   0.015 & 28.3094 $\pm$ 0.0028 & -      \\
F17 &   19.9681 $\pm$   0.0024  &       104     $\pm$   10      &       0.331   $\pm$   0.015 & 24.3584 $\pm$ 0.0029 & -      \\
F18 &   27.4616 $\pm$   0.0028  &       88      $\pm$   10      &       0.165   $\pm$   0.018 & 33.4995 $\pm$ 0.0034 & -      \\
F19 &   21.7361 $\pm$   0.0031  &       80      $\pm$   10      &       0.267   $\pm$   0.020 & 26.5151 $\pm$ 0.0037 & -      \\
F20 &   22.1513 $\pm$   0.0032  &       76      $\pm$   10      &       0.686   $\pm$   0.021 & 27.0216 $\pm$ 0.0039 & -      \\
F21 &   21.0256 $\pm$   0.0034  &       72      $\pm$   10      &       0.203   $\pm$   0.022 & 25.6484 $\pm$ 0.0041 & -      \\
F22 &   28.68628        $\pm$   0.0035  &       69      $\pm$   10      &       0.127   $\pm$   0.023 & 34.9934 $\pm$ 0.0042 & -      \\
F23 &   18.2134 $\pm$   0.0036  &       69      $\pm$   10      &       0.611   $\pm$   0.023 &       22.2179 $\pm$ 0.0044 & -        \\
F24 &   21.2856 $\pm$   0.0037  &       67      $\pm$   10      &       0.064   $\pm$   0.024 &       25.9656 $\pm$ 0.0045 & -        \\
F25 &   14.9793 $\pm$   0.0045  &       55      $\pm$   10      &       0.754   $\pm$   0.029 &       18.2727 $\pm$ 0.0054 & -        \\
F26 &   30.1311 $\pm$   0.0051  &       48      $\pm$   10      &       0.415   $\pm$   0.033 &       36.7559 $\pm$ 0.0062 & -        \\
F27 &   22.6975 $\pm$   0.0052  &       47      $\pm$   10      &       0.123   $\pm$   0.034 &       27.6879 $\pm$ 0.0063 & -        \\
F28 &   30.4605 $\pm$   0.0064  &       39      $\pm$   10      &       0.585   $\pm$   0.041 &       37.1577 $\pm$ 0.0078 & -        \\
F29 &   30.3283 $\pm$   0.0088  &       28      $\pm$   10      &       0.779   $\pm$   0.056 &       36.9965 $\pm$ 0.0107 & -        \\
                \hline
        \end{tabular}
\end{table*}

\subsection{Noise treatment}
\label{sec:noise}

Previous studies similar to this one, but based on other targets
\citep[see, e.g.,][]{Bourrier2019,Wong2020}, demonstrated that PDC
data showed additional time-correlated residual features in the
photometry that were absent from the SAP light curves. The authors
typically carried out their own detrending strategy over SAP data and
used these results as base for their analysis. In our case, the strong
intrinsic variability of the star manifests itself as stellar
pulsations, whose amplitudes are larger than the time-correlated
features mentioned before. Because the pulsations hinder a proper
analysis of the residual noise in our light curves, our approach in
this work is slightly different. First, we recall that the criteria
for considering a pulsation as detected are based on a very tight
constraint of the amplitude S/N, \mbox{AS/N $>$ 4}, which in turn is
determined around a box of 2 cd$^{-1}$, as detailed before. This
constraint is significantly tighter than, for example, the use of the
false-alarm probability in a periodogram to claim the detection of a
pulsation frequency. The detected pulsations are therefore robust,
despite the residual features of the PDC. Additionally, even though we
have characterized the pulsation spectrum of the host star as has
never been done before, there are residual pulsations in TESS
photometry. To take the effect of the residual features in the PDC
data into consideratio that we cannot clearly characterize, and the
residual pulsation features left after accounting for our derived 29
pulsation frequencies in both SAP and PDC light curves, we computed in
both cases the $\beta$ factor, as specified by \cite{Carter2009}, and
used the minimization of the $\beta$ factor as a tool to determine
which data set to use. To quantify to what extent TESS photometry is
affected by systematic noise, we computed residuals for the PDC and
SAP data by subtracting the primary transit light curves and the 29
pulsation frequencies. We then divided the residuals into M bins of
equal duration, with N equal to the number of data points per
bin. When the residuals are not affected by red noise, they probably
follow the expectation of independent random numbers,

\begin{equation}
  \sigma_N = \sigma_1 N^{-1/2}[M/(M-1)]^{1/2}\ ,
\end{equation}

\noindent where $\sigma_1^2$ is the sample variance of the unbinned
data and $\sigma_N^2$ is the sample variance (or RMS) of the binned
data, with the following expression:

\begin{equation}
  \sigma_N = \sqrt{\frac{1}{M}\sum_{i = 1}^{M}(<\hat{\mu_i}> - \hat{\mu_i})^2}.
\end{equation}

\begin{figure}[ht!]
    \centering
    \includegraphics[width=0.5\textwidth]{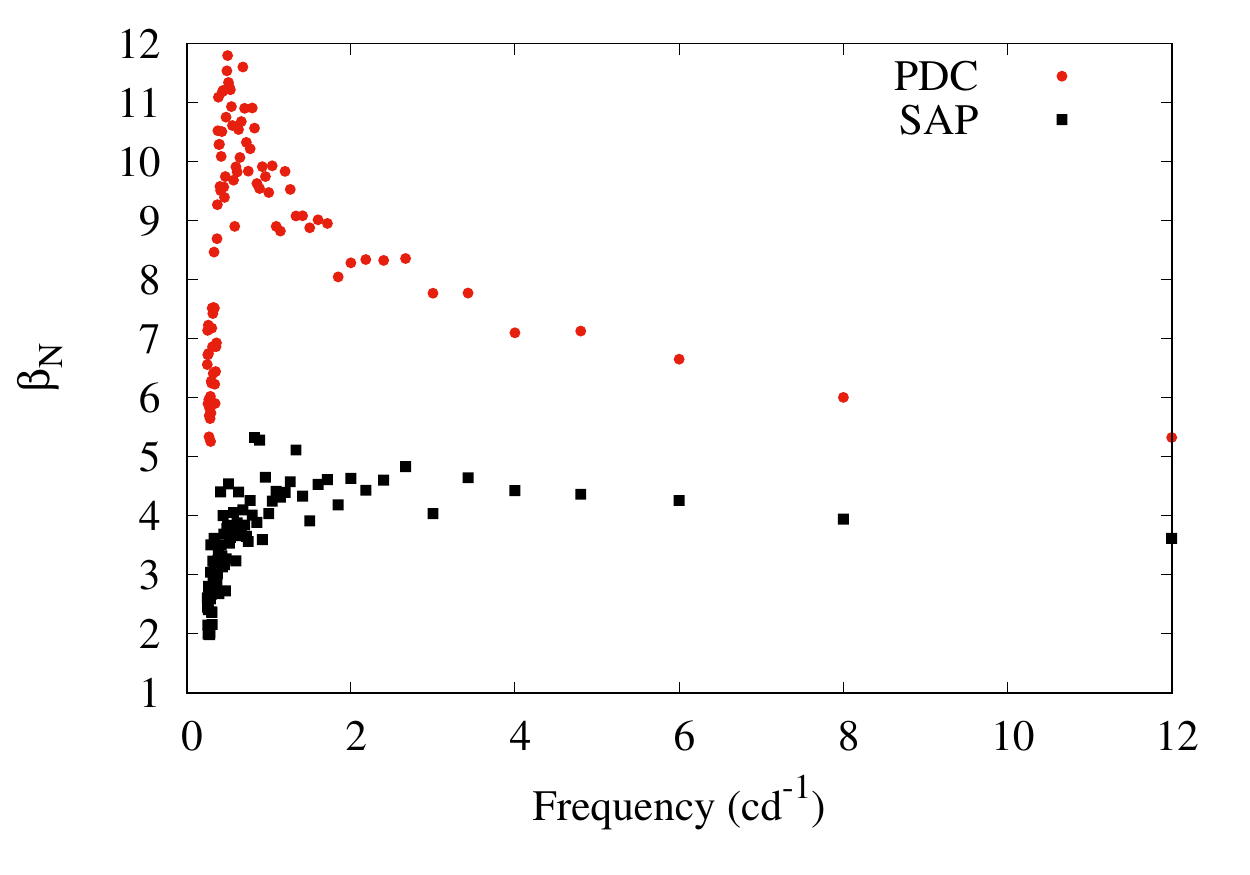}
    \caption{Correlated noise in the TESS photometric products. $\beta_N$ as a function of frequency (1/$\Delta$t) for the PDC (red circles) and SAP (black squares) data sets.}
    \label{fig:beta_factor}
\end{figure}

\noindent In the equation, $\hat{\mu_i}$ is the mean value of the
residuals per bin, and $<\hat{\mu_i}>$ is the mean value of the
means. In the presence of correlated noise, each $\sigma_N$ differs by
a factor of $\beta_N$ from their expectation. The $\beta$ factor, used
to enlarge the individual photometric errors, is an average of all
$\beta_N$ computed considering different timescales, $\Delta$t, that
are judged to be most important. In our case, the nature of the noise
and thus the relevant timescales differs drastically. To compute
$\beta_N$ , we therefore considered $\Delta$t between 2 hours
(systematics due to residual pulsations) and 4 days (systematics due
to PDM detrending), divided into steps of 1 hour, corresponding to a
total of 94 $\Delta$t. Figure~\ref{fig:beta_factor} shows our
resulting $\beta_N$ as a function of the frequency (1/$\Delta$t). In
both cases, the highest $\beta_N$ values are at the lower end of the
frequency range, where the TESS total time coverage does not allow us
to resolve the pulsation frequencies from the high noise.  $\beta_N$
increases drastically from SAP to PDC photometry by almost a factor of
three. The corresponding $\beta$ factors are computed from the average
of the $\beta_N$, with values \mbox{$\beta_{SAP}$ = 3.545}, and
\mbox{$\beta_{PDC}$ = 8.609}. This work is therefore based on SAP
data, where the individual error bars are increased by a factor of
3.545.

\subsection{Primary transit parameters from TESS light curves}
\label{sec:PT}

\cite{vonEssen2014} derived the transit parameters with and without
the intrinsic variability of the host star. After comparing the
derived parameters, we found no significant differences in the two
sets. The analysis of TESS data revealed no different results, which
were derived in the following way. First, we analyzed TESS data
including all the pulsation frequencies. We considered the data points
around $\pm$0.1 days centered on each mid-transit time, so that we had
enough off-transit data to normalize each transit. For the
normalization we used a second-order time-dependent polynomial that
was simultaneously fit to the transit model
\citep{MandelAgol2002}. The degree of the polynomial was chosen from a
prior analysis of the data, and as detrending functions, we tried a
first-, second-, and third-degree time-dependent polynomial. After
carrying out a simple least-squares fit between the primary transits
and the detrending times the transit model, we computed the Bayesian
information criterion (BIC) from the residuals, and analyzed which
polynomial systematically minimized the BIC. The fitting parameters
were the semimajor axis, a/R$_S$, the inclination in degrees, i, the
orbital period, P, the planet-to-star radius ratio, R$_P$/R$_S$, and
the mid-transit time of reference, T$_0$. As specified before, we used
a quadratic limb-darkening law. We considered the eccentricity fixed
and equal to zero \citep{Smith2011}. Our joined model includes
3$\times$TN + 5 parameters, where \mbox{TN = 16} corresponds to the
total number of transits, 3 to the number of coefficients for the
detrending polynomial, and 5 accounts for the primary transit
parameters previously mentioned. The transit light curves, along with
the derived best-fit transit model, are shown in
Figures~\ref{fig:transits1} and \ref{fig:transits2} . Subsequently, we
analyzed TESS data after removing all the pulsation frequencies listed
in Table~\ref{tab:pulsations}, specifically, dividing away from the
fluxes the summation of the 29 pulsation frequencies using the
following equation:

\begin{equation}
    \mathrm{PM(t) = \sum_{i=1}^{29}A_i \times \sin[2\pi(t\nu_i + \phi_i)]},\,
    \label{eq:PM}
\end{equation}

\noindent where A$_i$, $\nu_i$ and $\phi_i$ are the amplitudes,
frequencies, and phases listed in Table~\ref{tab:pulsations},
respectively. The number of fitting parameters and the degree of the
detrending function remained unchanged. All the primary transit light
curves cleaned of the 29 pulsation frequencies, along with the
best-fit transit model, are also shown in Figures~\ref{fig:transits1}
and \ref{fig:transits2}, vertically shifted down to allow for visual
comparison.  To derive the best-fit values for the model parameters
and their corresponding uncertainties, we used an MCMC approach, which
is implemented in the routines of
PyAstronomy\footnote{www.hs.uni-hamburg.de/DE/Ins/Per/Czesla/PyA/PyA/index.html}
\citep{Patil2010,Jones2001}. We iterated \mbox{1\ 000\ 000} times,
with a conservative burn-in of the first 20\% samples. For all the
parameters we considered uniform priors around $\pm$50\% of their
corresponding starting values. These were taken from
\cite{vonEssen2014} for the primary transit model, as specified in the
last column of Table~\ref{tab:transit_parameters}. Starting values for
each second-order time-dependent polynomial were derived from a simple
least-squares minimization. We computed the median and standard
deviation from the posterior distributions, and used these as our
best-fit values and uncertainties, given at 1$\sigma$ level. We
confirmed the convergence of the chains by visually inspecting each
one of them, and by dividing the chains into three subchains. In each
case, we computed the usual statistics, and we considered that a chain
converged when the derived parameters were consistent within their
uncertainties.

\begin{table*}[ht!]
    \centering
    \caption{\label{tab:transit_parameters} Best-fit transit
      parameters obtained from TESS photometry (this work), compared
      to those determined by \cite{vonEssen2014}, accounting for
      pulsations.}
    \begin{tabular}{l l l l}
    \hline\hline
    Parameter         &    This work                             &      This work                        &  \cite{vonEssen2014}   \\
                      & (not accounting for pulsations)   &  (accounting for pulsations)    &  (accounting for pulsations)  \\
    \hline
    a/R$_S$           & 3.605 $\pm$ 0.009                        &    3.614 $\pm$ 0.009                  &  3.68 $\pm$ 0.03 \\
    i ($^{\circ}$)    & 88.05 $\pm$ 0.28                         &    88.01 $\pm$ 0.28                   &  87.90 $\pm$ 0.93 \\
    R$_P$/R$_S$       & 0.10716 $\pm$ 0.00023                    &    0.10714 $\pm$ 0.00024              &  0.1046 $\pm$ 0.0006 \\
    P (days)    & 1.2198696 $\pm$ 4.2$\times$10$^{-6}$     & 1.2198681 $\pm$ 4.2$\times$10$^{-6}$  & 1.2198675 $\pm$ 1.5$\times$10$^{-6}$ \\
    T$_0$ (BJD$_\mathrm{TDB}$) & 2458792.63376 $\pm$ 0.00009     & 2458792.63403 $\pm$ 0.00009           &  2455507.5222 $\pm$ 0.0003 \\
    \hline
    \end{tabular}
\end{table*}

The ephemerides following from 23 days of TESS data, derived from our transit fitting accounting for the stellar pulsations, are the following:

\begin{center}
$\mathrm{T_0 = 2458792.63403 \pm 0.00009}$ BJD$_\mathrm{TDB}$

$\mathrm{Per = 1.2198681 \pm 4.2\times10^{-6}}$ days.
\end{center}

\begin{figure*}[ht!]
    \centering
    \includegraphics[width=\textwidth]{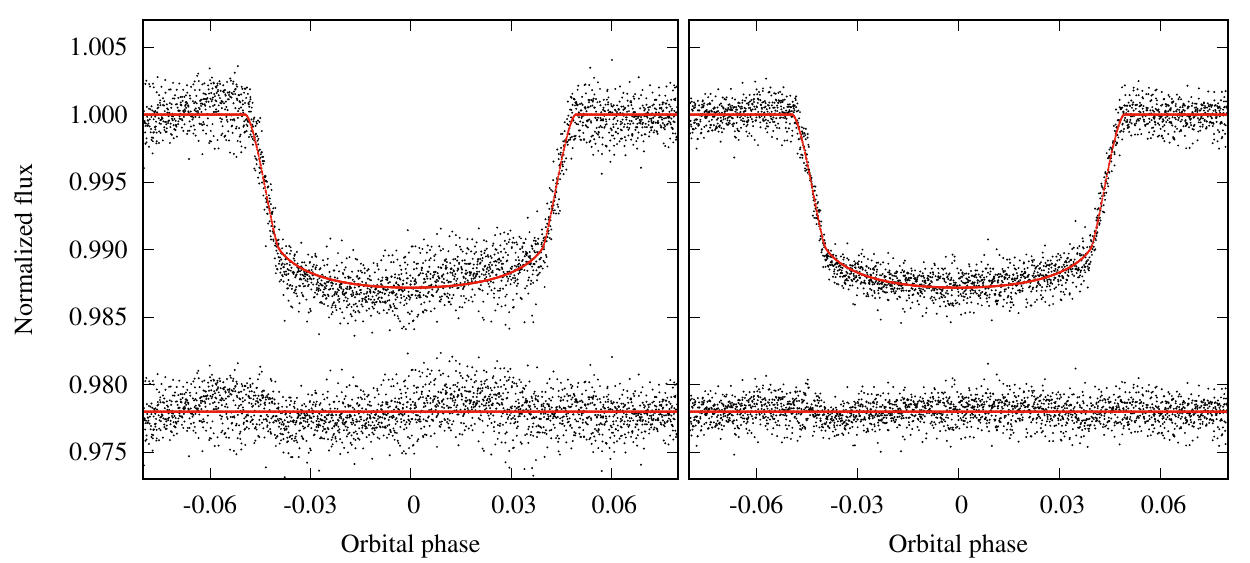}
    \caption{\label{fig:transits_folded} Primary transit light curves
      without (left) and with (right) stellar pulsations. Black points
      correspond to TESS data, and the red continuous line shows the
      derived best-fit transit model. Corresponding residuals are
      shown below each transit.}
\end{figure*}

Our derived period agrees within 1$\sigma$ with the period reported by \cite{Maciejewski2018}. The semimajor axis and inclination agree at the same level with the corresponding values reported in \cite{vonEssen2014}. The only exception is the planet-to-star radius ratio. Even though this value does not agree with the value reported by \cite{vonEssen2014}, it perfectly agrees with the aluminum oxide model extrapolated to TESS wavelengths \citep[see, e.g.,][]{vonEssen2019b,Welbanks2019}. This value also shows the accuracy on our treatment for third-light contribution. The phase-folded primary transit light curves are shown in Figure~\ref{fig:transits_folded}. The asymmetry in the transit shape is caused by the improper treatment of the pulsation frequencies. For the further analysis, we solely consider the transit parameters listed in the third column of Table~\ref{tab:transit_parameters}. 

\begin{figure}[ht!]
    \centering
    \includegraphics[width=0.5\textwidth]{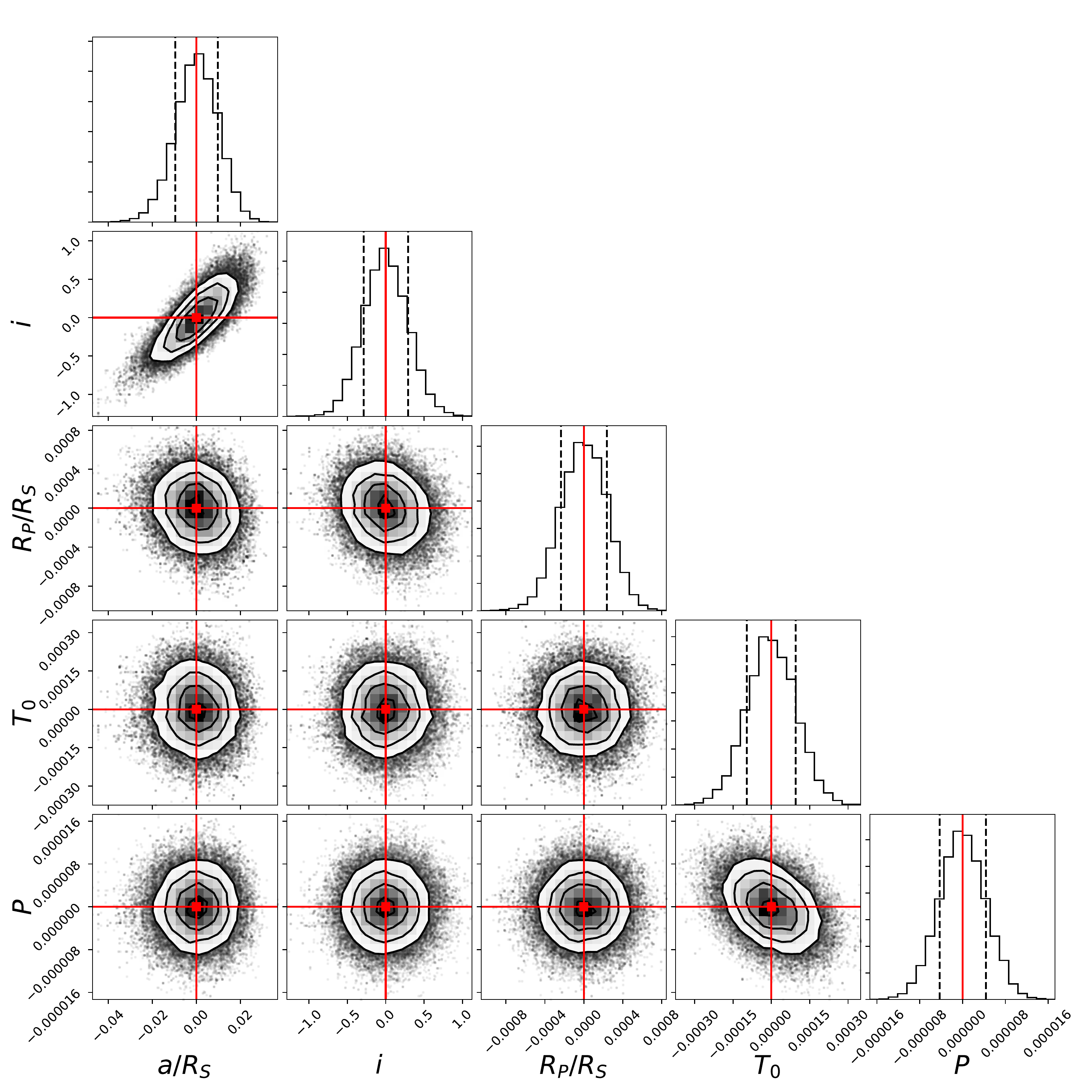}
    \caption{\label{fig:posterior_transits} Posterior distributions
      for the primary transit parameters. Gray to white contours
      indicate 1, 2, and 3$\sigma$ intervals. The red points
      correspond to the best-fit values. The chains were shifted to
      the best-fit values of the parameters specified in
      Table~\ref{tab:transit_parameters} to allow for a better visual
      inspection of the uncertainties.}
\end{figure}

The posterior distributions and the corresponding correlations between
parameters are presented in Fig.~\ref{fig:posterior_transits}. In
addition to the well-known correlation between a/R$_S$ and i, and to a
lesser extent, between P and T$_0$, the remaining parameters are
uncorrelated, with Pearson correlation values ranging between -0.05
and 0.04.

\subsection{Updated stellar and planetary parameters}
\label{sec:updated_params}

We redetermined the stellar and planetary radii and masses, taking
advantage of the newly available parallax from {\it Gaia\/} DR2
together with the available photometry from all-sky broadband
catalogs. We used the semi-empirical approach of measuring the stellar
spectral energy distribution (SED) described by \citet{Stassun:2017}
and \citet{Stassun:2016}.

\begin{figure}[!ht]
    \centering
    \includegraphics[width=6.7cm,angle=90,trim=70 80 90 90,clip]{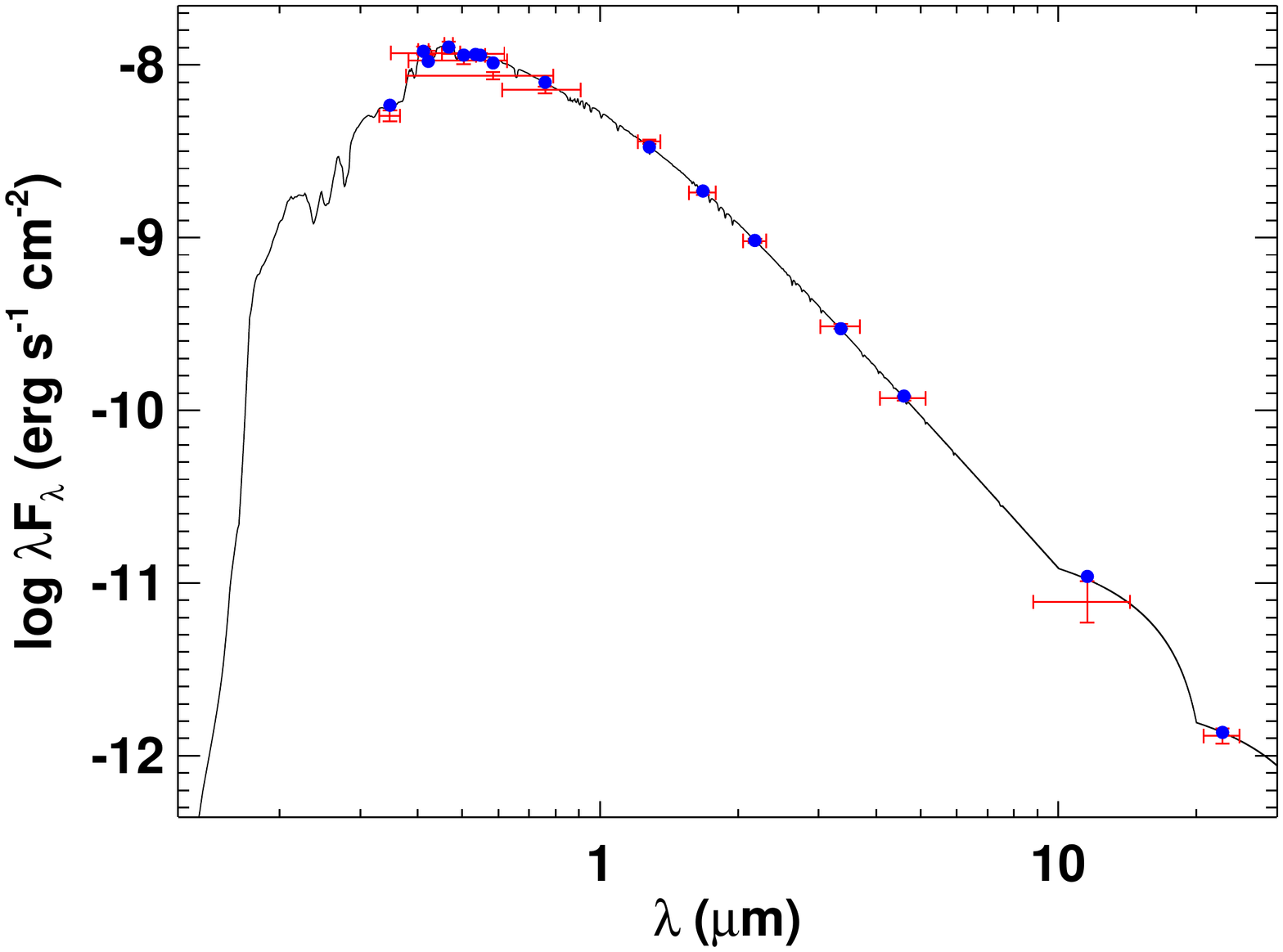}
\caption{SED of WASP-33. Red symbols represent the observed
  photometric measurements, where the horizontal bars represent the
  effective width of the passband. Blue symbols are the model fluxes
  from the best-fit Kurucz atmosphere model (black).
\label{fig:sed}}
\end{figure}

We pulled the $B_T V_T$ magnitudes from {\it Tycho-2}, the $uvby$
magnitudes from Str\"omgren $uvby$ magnitudes from
\citet{Paunzen:2015}, the $JHK_S$ magnitudes from {\it 2MASS}, the
W1--W4 magnitudes from {\it WISE}, and the $G G_{\rm BP} G_{\rm RP}$
magnitudes from {\it Gaia}. Together, the available photometry spans
the full stellar SED over the wavelength range 0.35--22~$\mu$m (see
Fig.~\ref{fig:sed}).

We performed a fit using Kurucz stellar atmosphere models, with the
priors on effective temperature ($T_{\rm eff}$), surface gravity
($\log g$), and metallicity ([Fe/H]) from the values reported in
Table~\ref{tab:third_light}. The remaining free parameter is the
extinction, $A_V$, which we limited to the maximum for the line of
sight from the dust maps of \citet{Schlegel:1998}. The resulting fit
(Fig.~\ref{fig:sed}) is very good, with a reduced $\chi^2$ of 2.7, and
best-fit $A_V = 0.04 \pm 0.04$. Integrating the (unreddened) model SED
gives the bolometric flux at Earth of $F_{\rm bol} = 1.455 \pm 0.051
\times 10^{-8}$~erg~s$^{-1}$~cm$^{-2}$. The $F_{\rm bol}$ and $T_{\rm
  eff}$ together with the {\it Gaia\/} DR2 parallax, adjusted by
$+0.08$~mas to account for the systematic offset reported by
\citet{Stassun:2018}, gives the stellar radius as $R_\star = 1.561 \pm
0.052$~$R_\odot$. Estimating the stellar mass from the empirical
relations of \citet{Torres:2010} gives $M_\star = 1.59 \pm
0.10$~$M_\odot$, which is consistent with that inferred from the
spectroscopic $\log g$ together with the radius ($1.77 \pm
0.82$~$M_\odot$).  Finally, the estimated mass together with the
radius gives the mean stellar density $\rho_\star = 0.59 \pm
0.07$~g~cm$^{-3}$.  With the updated value for the stellar radius and
the planet-to-star radii ratio derived fitting TESS photometry, we
report here the planetary radius in the TESS passband, \mbox{$R_P =
  1.627 \pm 0.054~R_J$}.

\subsection{Star-planet interaction}

Compared to \cite{vonEssen2014}, the quality of TESS data allowed us
to carry out a more thorough analysis of the pulsation spectrum of the
host star. Because we now count almost four times more pulsation
frequencies than before, we again investigated whether any of the
observed pulsations were induced by planetary tides over the star.

\begin{figure}[ht!]
    \centering
    \includegraphics[width=.5\textwidth]{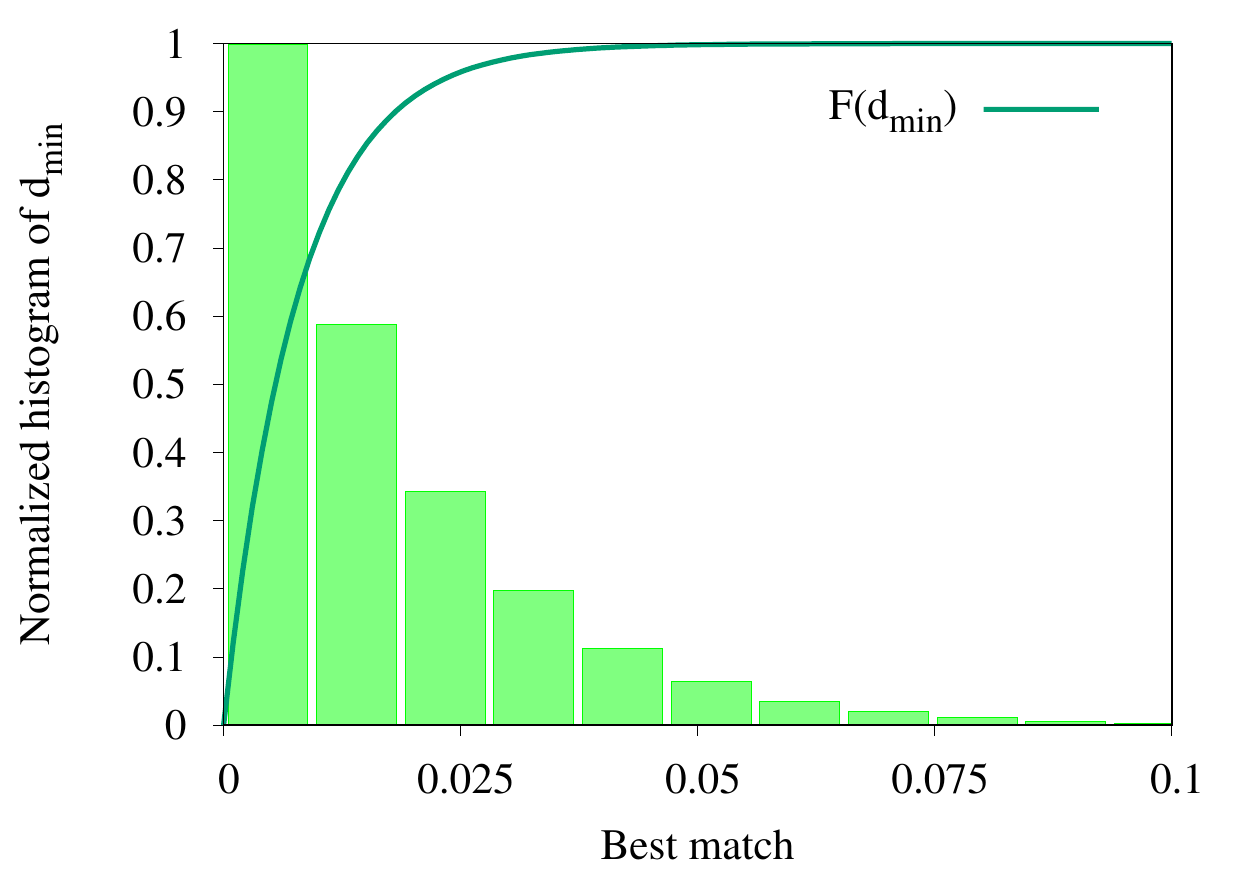}
    \caption{\label{fig:exp_decay} Normalized histogram for the
      best-match frequencies generated from 10$^6$ iterations,
      compared to F(d$_{min}$), plotted as the continuous line.}
\end{figure}

To begin with, we do not find a nonradial mode around $\sim$4~
cd$^{-1}$ as previously reported by \cite{CollierCameron2010}. In
addition to this, tidally excited modes can be manifested by the
commensurability between the orbital period of the planet and the
pulsation frequencies \citep[see, e.g.][]{Hambleton2013}. Similarly to
\cite{vonEssen2014}, we used our best-fit orbital period to express
the pulsation frequencies as cycles per orbital period
(cP$^{-1}$). These are given in the fifth column of
Table~\ref{tab:pulsations}. We found the closest commensurability to
be 36.99669 cP$^{-1}$, corresponding to the frequency \mbox{30.3283
  $\pm$ 0.0088 cd$^{-1}$}. The difference to its closest integer
number is equal to 0.00331. To assess whether this difference is
significant enough to pinpoint this pulsation as being triggered by
planetary tides, we carried out the same exercise as described in
\cite{vonEssen2014}. Briefly, we randomly generated 29 frequencies
between the lowest and highest detected frequencies that were in turn
derived from a uniform distribution. Then, we converted these
frequencies given in cd$^{-1}$ to cP$^{-1}$, and we selected the
frequency closest to an integer number. We then computed the
difference. We call this difference the "best match". After
1$\times$10$^6$ iterations, we computed the cumulative probability
distribution for the minimum distance from an integer frequency ratio,
$d_{min}$, as

\begin{equation}
F(d_{min}) = 1 - e^{-d_{min}/v}
,\end{equation}

\noindent where v is obtained from fitting our Monte Carlo results
with an exponential decay (see Figure~\ref{fig:exp_decay}). The
derived value is \mbox{v = 0.0078 $\pm$ 0.0007}. From this, we can
determine that the probability of finding at least one of the ratios
closer than 0.00331 c/per to an integer ratio, of 29 randomly produced
pulsation frequencies, is 35\%. This value is too high to
confidentially claim that this particular pulsation is induced by the
planet. It is therefore very unlikely that the system shows evidence
of star-planet interactions.

\section{Joint model}

\subsection{Fitting strategy}

Our model is the addition of five components that are explained in
detail below and that were fit to the unbinned TESS photometry in
simultaneous after the most prominent pulsations were removed (see
Section~\ref{sec:Puls_Impact}). Equivalently to the primary
transit-fitting approach, we used an MCMC to derive the best-fit
values of our model. In this case, we iterated \mbox{10\ 000} times,
with a burn-in of the first 2000 samples, after visually inspecting
the chains for convergence. The best-fit values for the parameters,
along with their corresponding 1$\sigma$ uncertainties, were derived
from the median and standard deviation of the posterior distributions,
regardless the use of uniform or Gaussian priors.

To reduce the number of fitting parameters, when a given parameter was
present in different model components, it was treated as equal by our
MCMC algorithm. For each model component we detail the model
parameters, differentiating those fit on their own and those treated
as equal by our MCMC algorithm. In all cases, the variable t
corresponds to the time provided by TESS. To ensure that our results
are not affected by our choice of priors, and to investigate whether
the planetary mass has any effect on our modeling, we carried out four
different fits in parallel. Model 1 (M1) has uniform priors on all the
parameters and considers the planetary mass as fixed. Model 2 (M2) has
Gaussian priors on all the parameters and considers the planetary mass
as fixed. Models 3 and 4 (M3 and M4) are similar to M1 and M2, with
the difference that the planetary mass is considered as a fitting
parameter. When we used uniform priors, we considered physically
reasonable ranges that always fulfilled a conservative $\pm$50\% of
each parameter, and when we used Gaussian priors, we either considered
the best-fit values and uncertainties derived in this work and
specified in the third column of Table~\ref{tab:transit_parameters} as
starting values and errors, or we used those from the literature when
not available from this work, for instance, the planetary mass. In
both cases, to be conservative, the uncertainties were enlarged a
factor of 3 when we used them as part of the Gaussian priors. At the
end of this section we provide the four sets of results, and report as
final values those corresponding to the smallest reduced
chi-squared. Particularly, for the case of the ellipsoidal variation
(Section~\ref{sec:EV}) and the Doppler beaming (Section~\ref{sec:DB}),
we followed the prescription provided and fully described by
\cite{Beatty2019}, which we do not repeat here.

\subsection{Primary transit model}

We modeled the WASP-33b primary transit model (TM(t)) as specified
before, using the model of \cite{MandelAgol2002} and our custom linear
and quadratic limb-darkening coefficients. The fitting parameters,
that is, the semimajor axis, a/R$_S$, the inclination in degrees, i,
the orbital period, P, the planet-to-star radius ratio, R$_P$/R$_S$,
and the mid-transit time of reference, T$_0$, are all parameters that
affect the different model components, and are therefore always
treated as equal. We recall that this joint model does not take into
account a second-order time-dependent polynomial to normalize the
light curves. Its use did not only account for the lack of the phase
curve model around primary transit, but also for imperfections in the
photometry (residual pulsations and other systematics present in the
SAP photometry, as previously discussed). In consequence, the
planet-to-star radius ratio will appear slightly enlarged to
compensate for the lack of normalization. As starting values and
standard deviation for the Gaussian priors, we used the values listed
in the third column of Table~\ref{tab:transit_parameters}. To be
conservative, the uncertainties were enlarged by a factor of 3.

\subsection{Secondary eclipse model}
\label{sec:SE_PC}

As secondary eclipse model we considered a scaled version of the
transit model given by \cite{MandelAgol2002}, with the linear and
quadratic limb-darkening coefficients set to zero. The contribution of
the WASP-33 companion and the orbital eccentricity were considered in
the same way as specified before. As described in
\cite{vonEssen2019a}, the secondary eclipse model, SE(t), is given by

\begin{equation}
\mathrm{SE(t)} = [\mathrm{TM(t)} - 1.] \times \mathrm{SF} + 1,\,
\end{equation}

\noindent where TM(t) corresponds to the primary transit model of
\cite{MandelAgol2002}, and SF corresponds to the scaling factor that
scales the transit to meet the secondary eclipse depth. From this
factor, the secondary eclipse depth is computed as
$(\mathrm{R_P/R_s})^2 \times \mathrm{SF}$, and its error is computed
from error propagation between the two. The fitting parameters are
those from the transit model and the scaling factor.

\subsection{Phase curve model}

As performed, for instance, by \cite{Cowan2008} and \cite{Zhang2018}
on the WASP-33 Spitzer photometry, we modeled the reflection of
starlight and thermal emission from the dayside of the planet, here
called planetary phase variability, PPV(t), as a series of first-order
expansions in sines and cosines:

\begin{equation}
\mathrm{PPV(t)} = c_0 + c_1 \times \mathrm{\sin(2\pi t/P)} + c_2 \times \mathrm{\cos(2\pi t/P)}.
\end{equation}

\noindent The fitting parameters are the offset, c$_0$, the amplitudes
of the sine and cosine, c$_1$ and c$_2$, respectively, and the orbital
period, treated as equal. This linear combination of sines and cosines
allows for a potential offset between the region of maximum brightness
and the substellar point.

\subsection{Ellipsoidal variation}
\label{sec:EV}

Gravitational effects of a close-in exoplanet on its host create a
distortion of the star that results in photometric variations with a
minimum that occurs twice per orbit: during primary transit, and
during secondary eclipse. This effect is called ellipsoidal variation,
and it mainly depends on the properties of the stellar surface, the
masses of planet and star, and their relative distances. The amplitude
of the ellipsoidal variation of WASP-33, A$_{EV}$, is given by

\begin{equation}
A_{EV} = \beta\frac{M_P}{M_S}\left(\frac{R_S}{a}\right)^3\end{equation}

\noindent \citep{Loeb2003}. Here, M$_P$ and M$_S$ correspond to the
planetary and stellar masses, respectively, a/R$_S$ is the semimajor
axis scaled to the stellar radius, and $\beta$ \citep{Morris1985}
depends on the limb- and gravity-darkening coefficients. For WASP-33A
and b, \mbox{A$_{EV}$ = 27.4~ppm}. The limb- and gravity-darkening
coefficients were taken from \cite{Claret2017}, interpolated to the
stellar parameters of WASP-33, and the mass of WASP-33b was taken from
\cite{Lehmann2015} with $M_P = 2.1 \pm 0.2$~M$_\mathrm{J}$. The
remaining parameters were taken from this work. For this model
component, the fitting parameters are those connected to TM(t),
treated as equal, and the planetary mass.

\subsection{Doppler beaming}
\label{sec:DB}

The planet and star both orbit their common barycenter. While doing
so, the star periodically moves toward and away from the
observer. This creates a variation in the brightness of the host star
that is in total synchronization with the frequency of the planet. The
amplitude of the Doppler beaming is given by

\begin{equation}
A_{DB} = (3 - \alpha)\frac{K_{RV}}{c},\,
\end{equation}

\noindent \citep{Rybicki1979}, where $\alpha$ is defined in
\cite{Loeb2003}, K$_{RV}$ is the radial velocity amplitude of the
planet, and c is the speed of light. For WASP-33, \mbox{A$_{DB}$ =
  2.7~ppm}. As in the case of the ellipsoidal variation, the fitting
parameters are those connected to TM(t) and the planetary mass, which
is treated as equal to the mass from the ellipsoidal variation.
 
\begin{figure}[ht!]
    \centering
    \includegraphics[width=.5\textwidth]{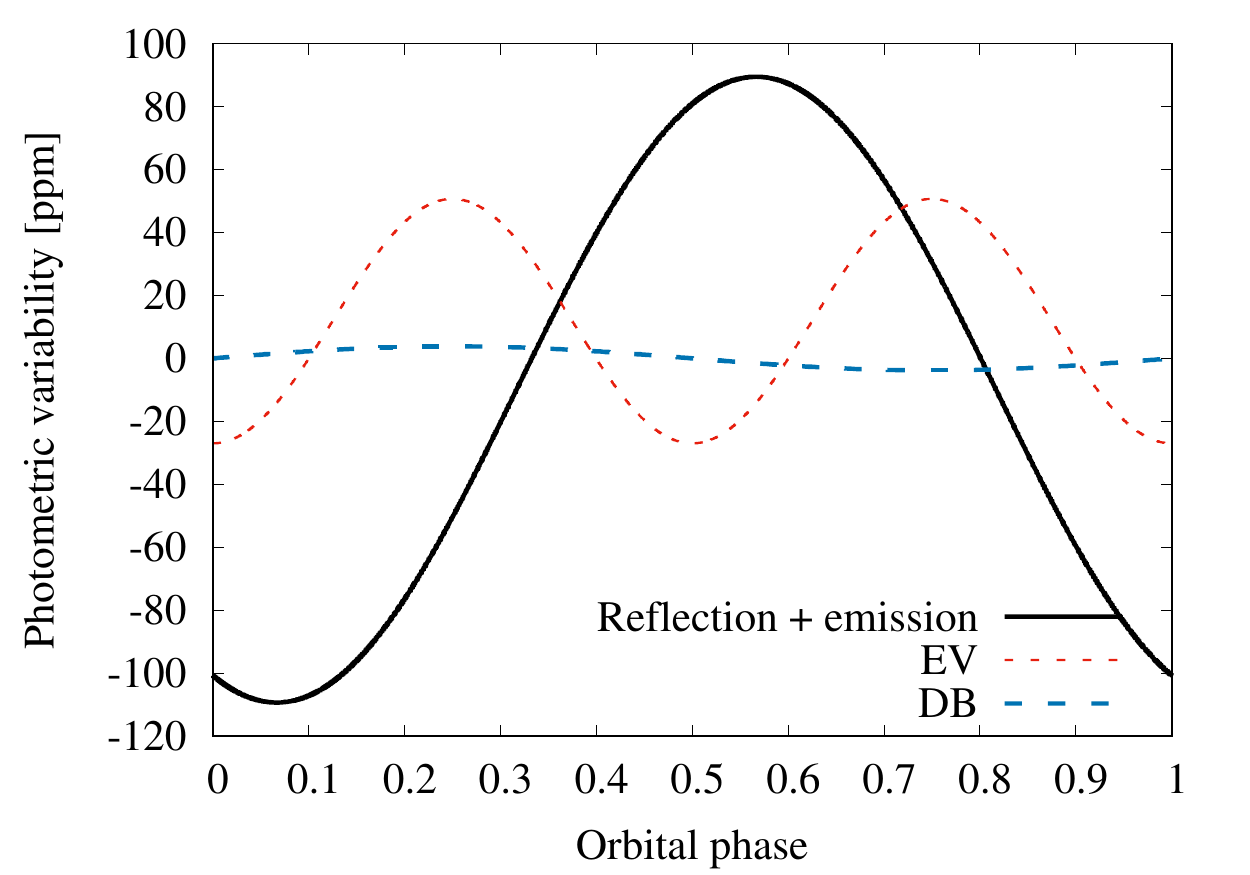}
    \caption{\label{fig:diff_models} Photometric variability in ppm
      given by reflection and emission shown as the black continuous
      line, the EV as the red dotted line, and the DB as the blue
      dashed line, as derived in this work. Values are placed around
      zero to allow for visual comparison.}
\end{figure}

Figure~\ref{fig:diff_models} shows the thermal emission and reflected
light, the ellipsoidal variation (EV), and the Doppler beaming (DB)
for WASP-33b, to allow for a comparison of the amplitudes of the
different effects. Even though the phase curve model is shown along
the whole orbital phase, in our joint fit it is turned off during
secondary eclipse because the planet is blocked by the star and
therefore does not produce any signal. As previously mentioned, in
Table~\ref{tab:exo_params} we provide four sets of results, with
uniform and Gaussian priors, and with and without fitting for the
planetary mass.

\subsection{Pulsation effect on the derived secondary eclipse depth and phase curve amplitude}
\label{sec:Puls_Impact}

As reported in Sect.~\ref{sec:PSpec}, 29 pulsations showed a
significant signal above the noise and were accordingly detected as
such. From Period04 we extracted the corresponding frequencies,
amplitudes, and phases, which were used to clean the data from
pulsations to recover the planetary signature. To do so, similarly to
\cite{vonEssen2014,vonEssen2015} , we considered as pulsation model
Eq.~\ref{eq:PM}. The pulsation amplitudes listed at the top of the
table are comparable to or even larger than the expected eclipse
depth. Instead of considering them as noise, we carried out a thorough
analysis of their effect over the planetary signature. As a
counterpart, several high-frequency pulsations reported in this work
have amplitudes smaller than 100~ppm, which means that they lie at the
limit or even below the TESS point-to-point scatter. In consequence,
these pulsations might be statistically irrelevant when they are
included as part of our model budget, which was also analyzed.

Using Period04, we computed pulsation-corrected light curves (PCLCs)
taking into account subgroups of pulsations, that is, the full set,
and those with the 25, 20, 15, 10, 9, 8, 7, 6, 5, 4, 3, 2, and 1
highest amplitudes. The difference in step served as a way to carry
out a more detailed investigation of the effect that the pulsation
frequencies with the highest amplitudes have. We ended up with 14
PCLCs, each one of them with a different "pulsation noise" level. We
prefer to use Period04 residuals because MCMC fits are
time-consuming. The former has been shown to deliver robust results
for pulsation frequency analysis.

To test whether the chosen number of pulsations affects the
determination of the physical properties of WASP-33b, we repeated the
same process for each of the 14 PCLCs. To speed the process up, we
subtracted the primary transit model as evaluated with the parameters
reported in Table~\ref{tab:transit_parameters}, plus the EV and the DB
that were evaluated considering the mass value given by
\cite{Lehmann2015}. Thus, we carried out an MCMC fit between the PCLCs
and the phase curve and secondary eclipse models specified before. In
each case we computed the best-fit parameters and their uncertainties,
along with the standard deviation of the residuals (PCLCs minus
best-fit SE(t) + PPV(t) model) and the \mbox{BIC = $\chi^2$ + k
  ln(N)}. For the BIC, $\chi^2$ was computed in the usual way, between
the PCLCs and the best-fit model. N corresponds to the total length of
the photometry, and k is the number of fitting parameters. For k we
considered the usual four parameters (SF, c$_0$, c$_1$ , and c$_2$),
plus 3$\times$PN, where PN is the number of pulsations considered in
each subgroup of PCLCs. The factor 3 accounts for each frequency,
amplitude, and phase.

\begin{figure}[ht!]
    \centering
    \includegraphics[width=.5\textwidth]{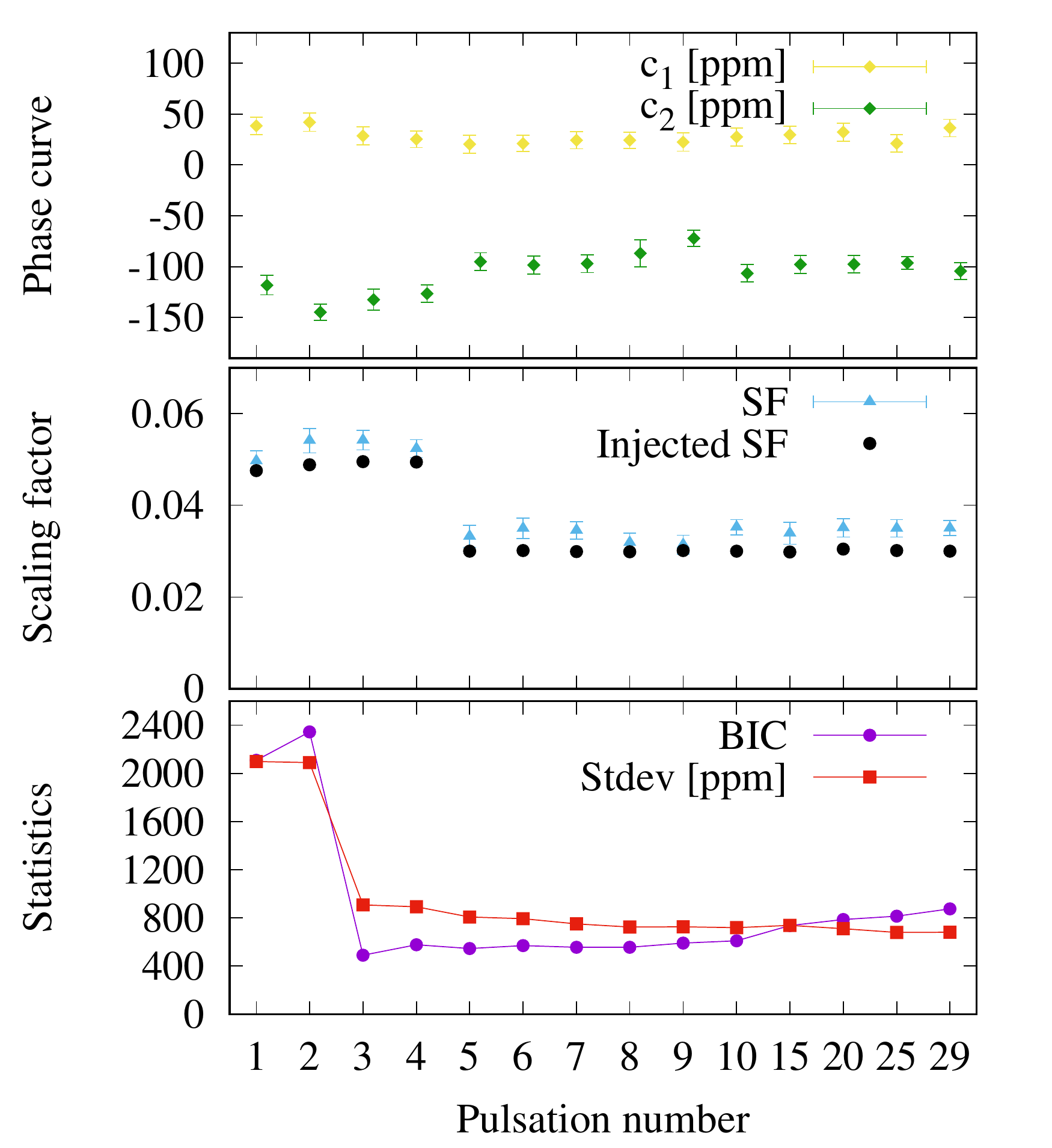}
    \caption{\label{fig:PFRs} Evolution of relevant parameters as a
      function of the number of pulsations considered to remove from
      the data. From top to bottom: the phase curve coefficients,
      c$_1$ and c$_2$, are shown as green points; the scaling factor
      and its injected counterpart are plotted as blue triangles and
      black circles, respectively, see text for details; and two
      statistics, the BIC as red circles and the standard deviation of
      the residual light curves as black squares, which was enlarged
      by a factor of 5 for better visualization.}
\end{figure}

Fig.~\ref{fig:PFRs} shows the evolution of the BIC, the standard
deviation of the residuals, and some of the fitted parameters,
specifically, SF, c$_1$ . and c$_2$, which were computed after the
primary transit, the ellipsoidal variation, and the DB were fit and
removed in a previous step. This means that only the phase curve and
the eclipse depth were isolated. These are given as a function of the
number of pulsations taken into consideration in the pulsation
model. The figure reveals two important aspects. First, a pulsation
model that includes the frequencies with the three highest amplitudes
favors the minimization of the BIC. However, there is a change by
almost 50\% in the scaling factor, when the first five frequencies
are included. To investigate if this large difference is caused by an
inadequate consideration of the pulsations of the host star, we
proceeded as follows. We created 14 light curves using as time stamps
those of TESS data, and as model, (SE(t) + PPV(t))$\times$PM(t). As
frequencies, amplitudes and phases we used those shown in
Table~\ref{tab:pulsations}, considering them in the same way as the
PCLCs were produced. As scaling factor we considered an arbitrary
value of 0.03. For c$_1$ and c$_2$, we took 50 and -120 ppm,
respectively. After the synthetic light curves were generated, we fit
them back with the phase curve and secondary eclipse models. The black
circles shown in the central panel of Fig.~\ref{fig:PFRs} reveal the
retrieved scaling factors for each of the synthetic light curves. As
expected, the recovered SFs follow the exact same behavior as those
that were obtained from TESS photometry. The observed jump is
therefore caused by an insufficient treatment of the pulsations of the
host star. It does not come as a surprise that adding F5 into the
pulsation modeling changes the parameters significantly because this
pulsation frequency is close to 3 cP$^{-1}$. From \mbox{PN = 5} and
onward, the derived SF is consistent within the errors. Among all
these, \mbox{PN = 5} corresponds to the smallest BIC value. In
addition to this, the difference between the two lowest BIC values
exceeds the required \mbox{$\Delta$BIC $< -6$} \citep{Kass1995}, which
is strong evidence in favor of the pulsation model formed solely by
the first five frequencies. The remaining pulsation frequencies have a
negligible effect on the planetary features, and are therefore ignored
for further analyses.

\section{Discussion and results}
\label{sec:discussion}

\subsection{Model parameters derived for WASP-33b}
\label{sec:params_W33_derived}

\begin{figure*}[ht!]
    \centering
    \includegraphics[width=.85\textwidth]{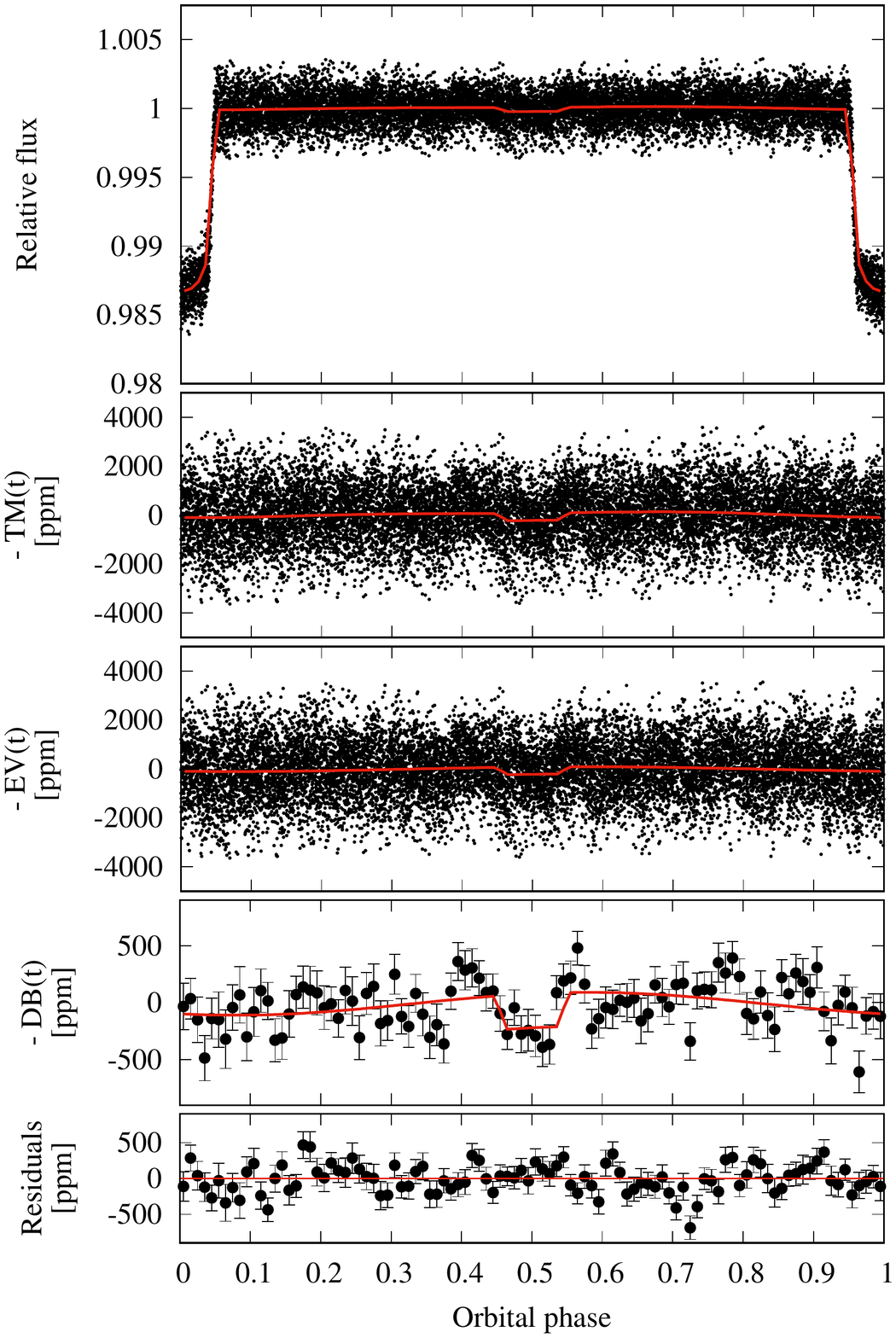}
    \caption{\label{fig:W33_final_model_data} From top to bottom:
      Phase-folded light curve of WASP-33 presented as black points
      showing the primary transit around phases 0,1 and the residual
      pulsations of the host star; relative flux of WASP-33 in parts
      per million (ppm) after removing the primary TM (second panel),
      the EV (third panel), and the DB (fourth panel). The last two
      panels show the data binned at \mbox{$\phi$ = 0.01} with the
      secondary eclipse plus phase curve at the top, and the residuals
      from the best-fit model at the bottom. Red lines always show the
      different components of the best-fit model.}
\end{figure*}

WASP-33b data, along with our best-fit joint model, are shown in
Fig.~\ref{fig:W33_final_model_data}. The most prominent first five
pulsation frequencies were removed from the light curve. The last two
panels are binned to \mbox{$\Delta\phi$ = 0.01} (equivalently,
$\Delta$t$\sim$17 minutes). At this cadence, the photometric precision
is 159 ppm. Panels 2 and 3 show no visible difference. With this we
emphasize that the amplitude of the EV and DB are significantly
smaller than that of the thermal emission and reflected light (see
Figure~\ref{fig:diff_models}). The best-fit values for the derived
parameters of the four model approaches, along with their
uncertainties computed in the usual way from their posterior
distributions, are listed in Table~\ref{tab:results_MCMC}.

We computed the amplitude of the phase curve as

\begin{equation}
    \mathrm{A(c_1, c_2) = \sqrt{c_1^2 + c_2^2}}\,,
\end{equation}

\noindent and we computed the eclipse depth as

\begin{equation}
   \mathrm{ED(SF, R_P/R_S) = SF \times (R_P/R_S)^2}.
\end{equation}

\noindent In both cases, the uncertainties were derived from error
propagation of the involved parameters.

\begin{table*}[ht!]  
    \centering
    \caption{\label{tab:results_MCMC} Best-fit values and 1$\sigma$
      uncertainties for the parameters accompanying, among others, the
      \cite{MandelAgol2002} model, a/R$_S$, i, R$_P$/R$_S$, P, and
      T$_0$, the phase curve, c$_0$, c$_1$ , and c$_2$, and the
      scaling factor of the secondary eclipse model, SF. In addition,
      the amplitude of the phase curve, A, the ED, and the offset
      between the region of maximum brightness and the substellar
      point, $\phi_\mathrm{off}$. Values are given, from left to
      right, for M1, M2, M3, and M4.}
    \begin{tabular}{l c c c c}
    \hline\hline
    Parameter             &   M1                    &   M2                &   M3                &  M4                 \\
    \hline
    a/R$_S$               &   3.601 $\pm$ 0.005     & 3.600 $\pm$ 0.005   & 3.603 $\pm$ 0.005   & 3.600 $\pm$ 0.005    \\
    i ($^{\circ}$)        &   88.53 $\pm$ 0.21      & 88.53 $\pm$ 0.22    & 88.52 $\pm$ 0.22    & 88.52 $\pm$ 0.23     \\
    R$_P$/R$_S$           &   0.1087 $\pm$ 0.0002   & 0.1087 $\pm$ 0.0002 & 0.1086 $\pm$ 0.0002 & 0.1087 $\pm$ 0.0002  \\
    P (days)        & 1.2198705 $\pm$ 3.7$\times$10$^{-6}$ & 1.2198706 $\pm$ 3.7$\times$10$^{-6}$ & 1.2198706 $\pm$ 3.9$\times$10$^{-6}$   & 1.2198704 $\pm$ 3.8$\times$10$^{-6}$  \\
    T$_0$ (BJD$_\mathrm{TDB}$) & 2458792.63405 $\pm$ 0.00006 & 2458792.63405 $\pm$ 0.00006 & 2458792.63405 $\pm$ 0.00005 & 2458792.63405 $\pm$ 0.00005  \\
    \hline
    c$_0$ (ppm)           &     1 $\pm$ 10          &  3 $\pm$ 10         & -56 $\pm$ 19        & -9 $\pm$ 12      \\
    c$_1$ (ppm)           &     45.4 $\pm$ 12.2     &  47.4 $\pm$ 10.7    & 52.8 $\pm$ 12.2     & 48.3 $\pm$ 12.1     \\
    c$_2$ (ppm)           &     87.7 $\pm$ 15.1     &  90.5 $\pm$ 14.9    & 87.4 $\pm$ 13.1     & 88.1 $\pm$ 13.4     \\
    SF                    &    0.0189 $\pm$ 0.0027  & 0.0270 $\pm$ 0.0031 & 0.0220 $\pm$ 0.0035 & 0.0258 $\pm$ 0.0030 \\
    \hline
    ED (ppm)              & 223.9 $\pm$ 31.0        & 320.4 $\pm$ 37.2    & 259.4 $\pm$ 41.0    & 305.8 $\pm$ 35.5    \\
    A (ppm)               & 98.7 $\pm$ 14.7         & 102.1 $\pm$ 14.1    & 102.1 $\pm$ 12.8    & 100.4 $\pm$ 13.1     \\
$\phi_\mathrm{off}$ ($^{\circ}$) & 27.3 $\pm$ 7.6   & 27.7 $\pm$ 6.6      & 31.1 $\pm$ 7.0      & 28.7 $\pm$ 7.1      \\
    M$_P$ (M$_\mathrm{J}$)& -                       & -                   & 5.7 $\pm$ 1.2       & 2.81 $\pm$ 0.53     \\
    \hline
    $\chi$                & 17273                   & 16101               & 17269               & 16100               \\
    BIC                   & 17359                   & 16187               & 17364               & 16192               \\
    $\chi^2_{red}$        & 1.2316                  & 1.1541              & 1.2314              & 1.1541              \\
    Degrees of freedom    & 9                       & 9                   & 10                  & 10                  \\
    \hline
    \end{tabular}
\end{table*}

Table~\ref{tab:results_MCMC} shows general consistency of the results,
not only when the different modeling approaches are compared, but also
when the transit parameters between this and those listed in
Table~\ref{tab:transit_parameters} are compared. An exception is the
ratio of the planet-to-star radius. As mentioned before, this
difference is caused by the lack of primary transit detrending in the
joint model because adding an unphysical detrending would absorb the
physics we wish to extract from the data. The photometry is so
extensive and so strongly deformed by the pulsations of the host star
that even the length of the off-transit data during the primary
transit-fitting stage had to be carefully chosen to avoid an effect on
the derived parameters. Thus, the transit parameters reported in this
work are those listed in the third column of
Table~\ref{tab:transit_parameters}. In any case, either choosing the
R$_P$/R$_S$ derived in Section~\ref{sec:PT} or the one obtained here
gives a fully consistent eclipse depth within its uncertainty. This
means that this slight transit parameter difference does not affect
our results for the phase curve and eclipse parameters. The model with
the smallest BIC/$\chi^2_{red}$ is M2, where the planetary mass is not
a fitting parameter. However, the $\Delta \mathrm{BIC}\,=\,-5$
compared to model M4 does not favor M2 significantly. To emphasize the
additional benefit of an independent mass measurement for WASP-33b
from the phase curve photometry, we discuss the results of M4 in the
remainder of this paper. Nonetheless, it is worth to mention that the
use of uniform and Gaussian priors on the mass of the planet returned
different results. This is probably because TESS data, even though
extremely rich, do not allow for a completely independent
determination of the planetary mass without previous knowledge of
it. Posterior probability distributions for the fitted parameters are
shown in Fig.~\ref{fig:posteriors_MCMC} for M4 alone because all
posteriors look alike.

\subsection{Physical parameters derived from these observations}
\label{sec:params_W33_derived_2}

Following the prescription given by \cite{Cowan2011}, from our derived
parameters we computed the Bond albedo,

\begin{equation}
    A_B = 1 - \frac{5 T_n^4 + 3 T_d^4}{2T_o^4}\,,
\end{equation}

\noindent and the heat redistribution efficiency,

\begin{equation}
    \epsilon = \frac{8}{5 + 3 (T_d/T_n)^4}
,\end{equation}

\noindent where $T_d$ and $T_n$ correspond to the temperature of the
dayside and the nightside, respectively. To compute the dayside
temperature, we made use of the secondary eclipse depth divided by the
primary transit depth because this is a direct measure of the ratio of
the planetary dayside intensity to the stellar intensity,
\mbox{$\psi(\lambda)_{day}$ = ED/(R$_P$/R$_S$)$^2$ = SF}. Equivalently
to this, to compute the nightside temperature, we used the difference
between the secondary eclipse depth and the phase variation amplitude,
combined with the offset between the region of maximum brightness and
the substellar point, \mbox{$\psi(\lambda)_{night}$ = ED -
  2A$\times$cos($\phi_\mathrm{off}$)}. We then computed the brightness
temperature of the planet following the prescription given in
\cite{Cowan2011},

\begin{equation}
T_b(\lambda) = \frac{hc}{\lambda k}\left[log\left(1 + \frac{e^{hc/\lambda k T*_b(\lambda)}-1}{\psi(\lambda)}\right)\right]^{-1} \, .
\end{equation}

\noindent Here, T$*_b(\lambda)$ is the stellar brightness temperature
within the TESS passband, equal to \mbox{7337 K}, and h, c, and k are the
Planck constant, the speed of light, and the Boltzmann constant,
respectively. Table~\ref{tab:exo_params} summarizes the values derived
in this work, compared to those computed by \cite{Zhang2018}. Results
for all modeling approaches are provided to show that our analysis is
reliable.

\begin{table*}[ht!]
    \centering
    \caption{\label{tab:exo_params} Bond albedo, recirculation efficiency, and day- and nightside brightness temperatures for WASP-33b within the TESS passband derived from M1, M2, M3, and M4. The last column shows the averaged values reported by \cite{Zhang2018} from Spitzer photometry.}
    \begin{tabular}{l c c c c c}
    \hline\hline
    Parameter          & M1                 & M2                 &  M3                &  M4                & \citep{Zhang2018}    \\
    \hline
    T$_{day}$ (K)      &  2881 $\pm$ 63     &  3037 $\pm$ 60     &  2934 $\pm$ 75     &  3014 $\pm$ 60     & 3144 $\pm$ 114 \\
    T$_{night}$ (K)    &  1487 $\pm$ 95     &  1617 $\pm$ 39     &  1541 $\pm$ 73     &  1605 $\pm$ 45     & 1757 $\pm$ 88 \\
    A$_B$              &  0.473 $\pm$ 0.046 &  0.351 $\pm$ 0.049 &  0.434 $\pm$ 0.057 &  0.369 $\pm$ 0.050 & 0.25$^{+0.09}_{-0.10}$ \\
    $\epsilon$         &  0.168 $\pm$ 0.030 &  0.189 $\pm$ 0.013 &  0.180 $\pm$ 0.021 &  0.189 $\pm$ 0.014 & 0.22$^{+0.05}_{-0.04}$ \\
    \hline
    \end{tabular}
\end{table*} 

\begin{figure}[ht!]
    \centering
    \includegraphics[width=.5\textwidth]{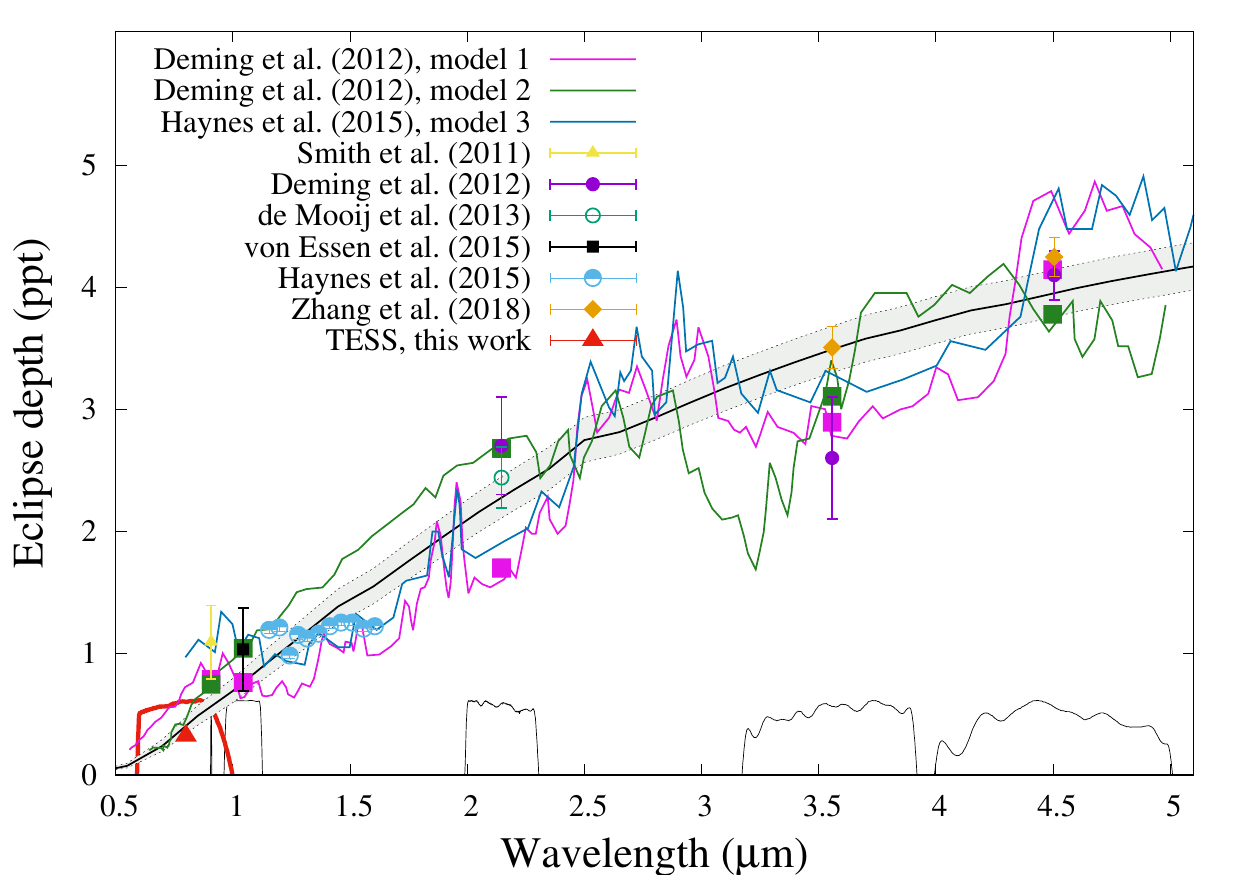}
    \caption{\label{fig:sec_W33} Eclipse depth, in ppt, as a function
      of wavelength. The red triangle corresponds to the TESS value,
      reported in this work. Error bars are included in the size of
      the point. The gray area shows 1$\sigma$ contour of the
      equilibrium temperature of WASP-33b. Literature measurements
      come from \cite{Smith2011} (triangle in yellow)
      \cite{Deming2012} (circles in violet), \cite{deMooij2013} (empty
      circle in green) \cite{vonEssen2015} (square in black),
      \cite{Haynes2015} (half-filled circles in blue),
      \cite{Zhang2018} (diamonds in yellow). Transmission responses
      are plotted as black continuous lines, with the exception of
      TESS, which is highlighted in red. Green and pink squares show
      \cite{Deming2012}'s models binned to their respective filters.}
\end{figure}

We obtained the uncertainties on these parameters using the posterior
probability distributions for the values c$_1$, c$_2$ , and SF. In
detail, for each of the 8000 MCMC iterations, we computed the Bond
albedo, recirculation efficiency, and day- and nightside brightness
temperature. In this way, the values reported in
Table~\ref{tab:exo_params} were obtained from their median and
standard deviation. Our values are fully consistent with those
reported by \cite{Zhang2018}. Fig.~\ref{fig:sec_W33} shows our derived
eclipse depth compared to literature measurements. Models 1 and 2 were
taken from \cite{Deming2012} and correspond to a solar composition
model with an inverted temperature in pink and a carbon-rich
noninverted model in green, respectively. Model 3 was taken from
\cite{Haynes2015} and reflects their best-fit inversion model with
TiO. To determine the effective temperature of WASP-33b, we fit all
the literature values to synthetic eclipse depths. These were created
by integrating the flux ratio between a blackbody of different
temperatures (the planet) and the PHOENIX intensities specified before
(the star). From $\chi^2$ minimization we obtained \mbox{T$_{eff,W33}$
  = 3105 $\pm$ 95 K}. The error on the temperature was computed
considering \mbox{$\Delta\chi^2$ = 1}.

When we fit model M4, the TESS phase curve also allowed for an
independent mass measurement of WASP-33b by a fit for the photometric
variations due to gravitational interactions between planet and host
star. We derive a value of $M_{P}=2.85 \pm 0.51$~M$_\mathrm{J}$, which
is consistent with the value of \cite{Lehmann2015} estimated by radial
velocity measurements within $1.5\,\sigma$. A good agreement between
photometric and radial velocity mass measurement was also achieved in
the TESS full-orbit phase curve of WASP-18b \citep{Shporer2019}, for
example. However, there are examples in the literature of discrepant
mass values of these independent methods that might be related to an
insufficient knowledge of the host star. We refer to the extensive
discussion of this problem in \cite{Shporer2017}.

\subsection{WASP-33b in context}
\label{sec:context}

Theoretical calculations predicted an energy transport by winds from
the dayside to the nightside on hot Jupiters
\citep{Showman2002,PerezBecker2013}. Early Spitzer phase curve
observations provided observational indications in line with these
predictions by measuring an eastward offset between the hottest
measured longitude and the most strongly irradiated longitude at the
substellar point \citep[e.g.,][]{Knutson2007,Knutson2012}.  With
increasing number of available Spitzer data, a trend was revealed of a
less pronounced phase offset and an increasing temperature day-night
contrast with increasing stellar insolation of the planet. It was
understood as a reduced efficiency of the heat transport for hotter
planets, caused by the balance between advective and radiative
cooling, and drag forces decelerating the advection
\citep[e.g.,][]{Komacek2017}. Surprisingly, two more recent
observations of the ultra-hot Jupiters WASP-103b and WASP-33b showed a
comparably weakened temperature contrast between day and night sides
\citep{Kreidberg2018,Zhang2018}. \cite{Bell2018} and \cite{Tan2019}
explained this effect by taking into account the dissociation and
recombination of hydrogen molecules, which plays a role only for the
very hottest known exoplanets. Our derived values for day and night
brightness temperature, and subsequently $\epsilon$, agree very well
with the results of \cite{Zhang2018} for WASP-33b, thus they
strengthen the indication of a turn-around in day-night temperature
contrast for the most insolated ultra-hot Jupiters. The TESS phase
curve analysis of the hottest known gas giant, KELT-9b, by
\cite{Wong2019b} also confirm this result. Associated with the same
phenomenon, \cite{Keating2019} described an increase of the nightside
temperature with increasing stellar insolation, including the
\cite{Zhang2018} temperature value of WASP-33b. Our result of this
work is in line with this correlation.

The TESS light curve revealed a westward phase offset for WASP-33b,
that is, the maximum in brightness occurred after the secondary
eclipse. Westward phase offsets like this have been spotted in
hot-Jupiter systems before. At NIR wavelengths, Spitzer 4.5 $\mu$m
revealed an offset of 21 $\pm$ 4 degrees in CoRoT-2b, which in turn
was not observed at 3.6 $\mu$m \citep{Dang2018}. To explain their
observational evidences, the authors offered three possible scenarios:
nonsynchronous rotation, magnetic effects, or eastern clouds. Even
rarer results were revealed by {\em Kepler} photometry in HAT-P-7b
because the observed phase offsets appeared to change in time
\citep{Armstrong2016}. Assuming these changes were not caused by
systematic noise over the data, \cite{Rogers2017} explained them by
magnetic interactions that might create phase offsets with periodic
changes in position. For a hot Jupiter of more moderate temperature,
Kepler-7b, the Kepler phase curves probe reflected light, therefore
the measured westward offset might be caused by a cloudy western
hemisphere \citep{Demory2013}. In addition to this work, another
ultra-hot Jupiter showing a westward offset of $\sim$14 degrees is
WASP-12b \citep{Bell2019}. A recent collection of phase offset
measurements are given in \cite{Keating2020}, who focused only on
Spitzer data at 3.6 and 4.5 $\mu$m. Their figure 7 reveals a wide
spread, with a clear preference for eastward offsets. However, about
six targets present westward offsets, and of this sample, two present
both eastward and westward phase offsets. For WASP-33b, the effective
temperature is above 3000 K. This means that the TESS wavelengths must
be probing mostly thermal emission and to a lesser extent, reflected
light. The observed westward offset might therefore be explained by
the three hypotheses presented by \cite{Dang2018} for
CoRoT-2b. However, in order to consolidate the observed eastward
offset by \cite{Zhang2018} in Spitzer 4.5 and 3.6 $\mu$m with our
detected westward offset in TESS light, time-variable clouds or
magnetic fields would be the best explanations. All these mixed
results might indicate that the atmospheric dynamics of hot and
ultra-hot Jupiters are more complex than we thought.

We add an additional word of caution. In an earlier stage of our data
analysis of this work, we carried out fits of the different model
components in independent steps, rather than simultaneously, as
presented here. The resulting phase offset we obtained was eastward
and consistent with the offset detected by \cite{Zhang2018}. In
consequence, we might consider that phase offsets are prone to details
in the analysis or systematics in the data. The reliable extraction of
the phase offset from photometric data might be more challenging than
we thought. Because the WASP-33 photometry is affected by intrinsic
variability of the host star, we cannot reject a nonastrophysical
origin of the westward phase offset.

The consistency in the measured dayside temperature between the
optical phase curve analyzed in this work and the NIR phase curves of
\cite{Zhang2018} indicate that the thermal emission of the WASP-33b
dayside is similar to a blackbody, potentially caused by the continuum
opacity of the hydrogen anion H$^-$
\citep{Arcangeli2018,Kitzmann2018}. However, the phase offset measured
in this work of $\phi_\mathrm{off}$ = $28.7 \pm 7.1$~degrees deviates
significantly from the negative (eastward) offset measured by
\cite{Zhang2018} in the NIR. If considered as a real astrophysical
phenomenon in the planetary atmosphere, a future 3D general
circulation modeling \citep[e.g.,][]{Kreidberg2018,Arcangeli2019}
might shed light on the underlying physical conditions.

Ultra-hot Jupiters such as WASP-33b are expected to be cloud-free on
their daysides because the atmospheres are too hot for condensates to
form, see \cite{Wakeford2017}. The lack of clouds is expected to
manifest itself as a low ability to reflect star light, thus a low
value of geometric albedo. The amount of light reflected off the
planet is included in the secondary eclipse depth, but it is merged
with the light that is thermally emitted by the planet itself. To
isolate and estimate the component of reflected light, we followed the
approach of \cite{Mallonn2019} and approximated the thermal emission
by a blackbody radiation of a temperature estimated from previous NIR
measurements. \cite{Pass2019} used Gaussian process regression to
derive a blackbody temperature of 3108~K for WASP-33b based on the
Hubble Space Telescope Wide Field Camera 3 (HST/WFC3) secondary
eclipse depths of \cite{Haynes2015} and the Spitzer results of
\cite{Zhang2018}. The stellar effective temperature is 7430~K
\citep{CollierCameron2010}.

By Equation~3 of \cite{Mallonn2019}, we estimate a geometric albedo of
$-0.04 \pm 0.04$, thus the 3\,$\sigma$ upper limit corresponds to
0.08. This low value of the geometric albedo is in line with the
theoretical expectation of no reflection because the temperature on
the dayside is too hot to form condensates. It also agrees with the
generally rather low measured optical values of other hot Jupiters by
the Kepler satellite \citep[e.g.,][]{Esteves2015,Angerhausen2015}, and
the ground-based z' band upper limits derived by
\cite{Mallonn2019}. However, the low geometric albedo contrasts with
the substantial value of the Bond albedo given in
Table~\ref{tab:exo_params}. Future secondary eclipse measurements near
the wavelength of stellar peak emission at 400~nm might shed light on
this issue. The null-detection of the geometric albedo proves that the
TESS phase curve of WASP-33b is dominated by thermal emission compared
to reflected light. The opposite has recently been found in the TESS
phase curve of WASP-19b \citep{Wong2020}, showing a significantly
nonzero albedo and dominating reflected light.

The approximation of the planetary thermal emission by a blackbody is
valid for the derivation of an upper limit of the geometric albedo
because more sophisticated emission modeling mostly points toward even
higher thermal flux in the optical for ultra-hot Jupiters
\citep{Haynes2015,Evans2019,Bourrier2019,Daylan2019}. Higher thermal
flux would translate into lower reflected light for a given value of
the eclipse depth, therefore an upper limit of the reflected light
component remains unaffected \citep{Mallonn2019}. We note that in the
case of TiO absorption at optical wavelengths, suggested by
\cite{Haynes2015} and \cite{Nugroho2017}, the optical eclipse depth is
predicted to be deeper than measured in this work, with a value of
$\sim$\,1000~ppm in the red part of the TESS bandpass \citep[see
  Figure~5 in][]{Haynes2015}. Hence, the TESS secondary eclipse depth
seems to disfavor the best-fit model of \cite{Haynes2015} that
includes TiO and a temperature inversion. However, because this
indication is based on multiple individual publications of secondary
eclipse depth in different wavelength regions, we suggest a
homogeneous reanalysis of these data sets before drawing a clear
conclusion, which is beyond scope of this work. Our suggestion is
strengthened by our result in Section~\ref{sec:Puls_Impact} that the
eclipse depth is affected by the number of pulsations included in the
modeling. All previous work on secondary eclipses has treated the
pulsations in a different way, thus a homogeneous reanalysis appears
useful.

\section{Conclusion}
\label{sec:conclusion}

We presented the first optical phase curve and secondary eclipse
observations of WASP-33b, obtained from analyzing 23 days of TESS
photometry. Based on the secondary eclipse depth, $ED = 305.8 \pm
35.5$~ppm, the amplitude of the phase curve, $A = 100.4 \pm 13.1$~ppm,
and the offset between the region of maximum brightness and the
substellar point of \mbox{28.7 $\pm$ 7.1} degrees, we used a simple
model to derive the brightness temperatures, $T_{day} = 3014 \pm 60$~K
and $T_{night} = 1605 \pm 45$~K, Bond albedo $A_B = 0.369 \pm 0.050,$
and recirculation efficiency, $\epsilon = 0.189 \pm 0.014$. While the
low geometric albedo of below 0.08 ($3\,\sigma$ upper limit) is
consistent with that of other hot Jupiters, the rather high
recirculation efficiency is consistent with previous WASP-33b studies
at NIR wavelengths and indicates the possibility of the dissociation
and recombination of hydrogen molecules in the atmospheres of
ultra-hot Jupiters. Additionally, the high photometric precision in
the phase-folded TESS data allowed for a mass measurement by the
photometric variations caused by gravitational interactions. This mass
agrees well with the literature value obtained from radial velocity
measurements. Because of the nature of the continuous TESS monitoring
of WASP-33, we characterized the pulsation spectrum of the host star,
finding 29 peaks with an AS/N higher than or equal to 4, instead of
the 8 known so far. The newly unveiled low-frequency range of the star
revealed two frequencies lower than 3 cd$^{-1}$ that are consistent
with gravity modes as observed in $\gamma$\,Doradus stars, making
WASP-33 a $\gamma$\,Doradus-$\delta$\,Scuti hybrid candidate. However,
more data are required to confirm this candidacy. Paying special
attention in the way the pulsation frequencies are considered while
determining planetary parameters, we find that using the minimization
of the BIC to quantify the amount of pulsation frequencies to be
considered in our model alone does not provide correct planetary
parameters. Special care has to be taken with pulsations with the
highest amplitude. Future detailed asteroseismic analyses of WASP-33
will help to improve the stellar parameters and better understand
possible star-planet interactions (or the lack thereof).

\section*{Acknowledgements}

We thank the referee for an outstanding quick reply, given the times,
and for contributing to significantly improve our work. CvE and GT
acknowledge support from the European Social Fund (project
No. 09.3.3-LMT-K-712-01-0103) under grant agreement with the
Lithuanian Science Council (LMTLT). Funding for the Stellar
Astrophysics Centre is provided by The Danish National Research
Foundation (Grant agreement no.: DNRF106). This work was supported by
a research grant (00028173) from VILLUM FONDEN.

\bibliographystyle{aa}
\bibliography{vonEssenC}

\begin{appendix}

\section{Primary transit photometry of WASP-33b}

\begin{figure*}[ht!]
    \centering
    \includegraphics[width=0.9\textwidth]{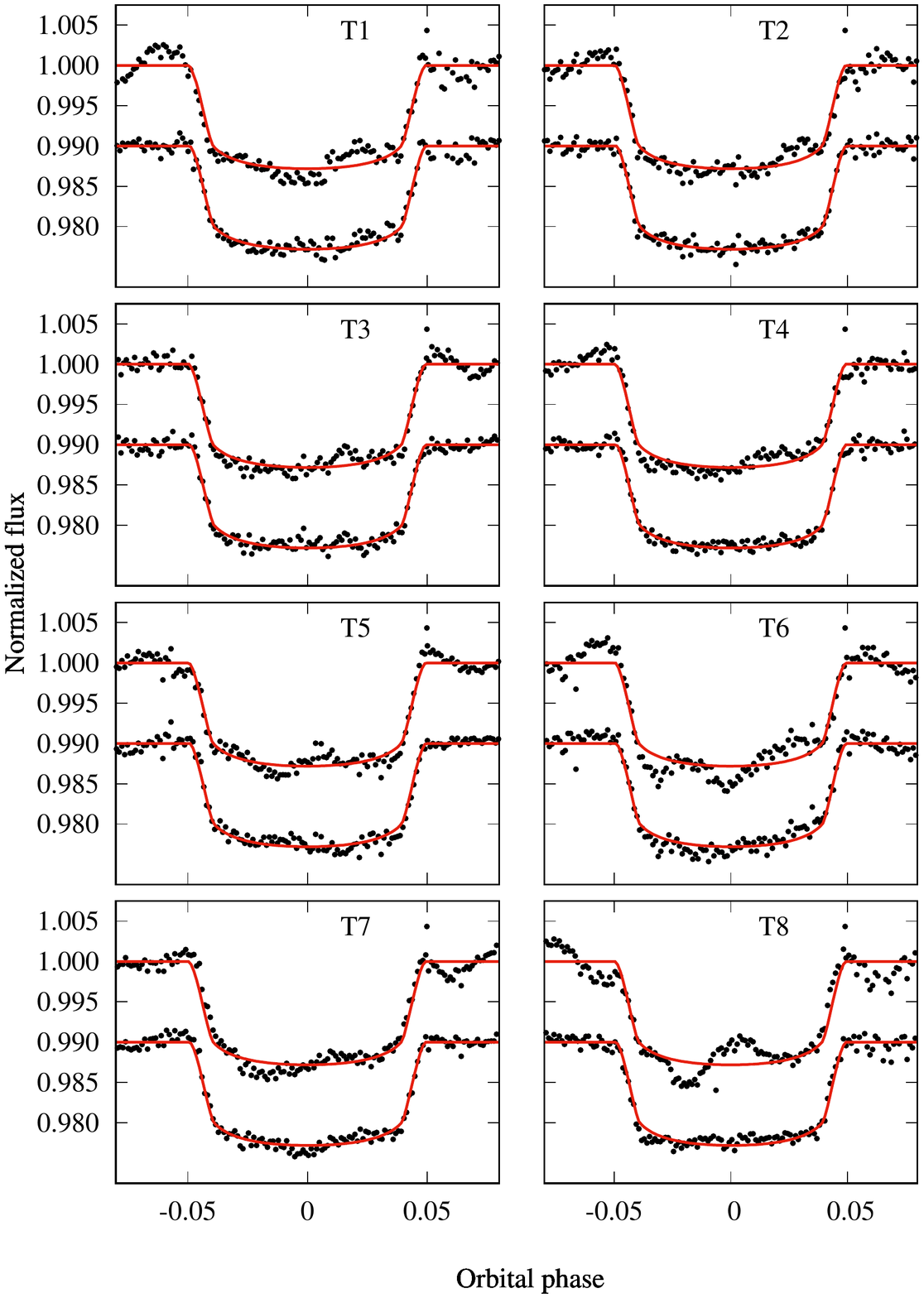}
    \caption{\label{fig:transits1} Normalized flux as a function of orbital phase for the primary transits of WASP-33b observed by TESS. From top to bottom and left to right time evolves. TESS observations are in black points, and the best-fit model in red continuous line. Individual error bars are not plotted to better visualize the effect of the pulsations over the photometry.}
\end{figure*}

\begin{figure*}[ht!]
    \centering
    \includegraphics[width=0.9\textwidth]{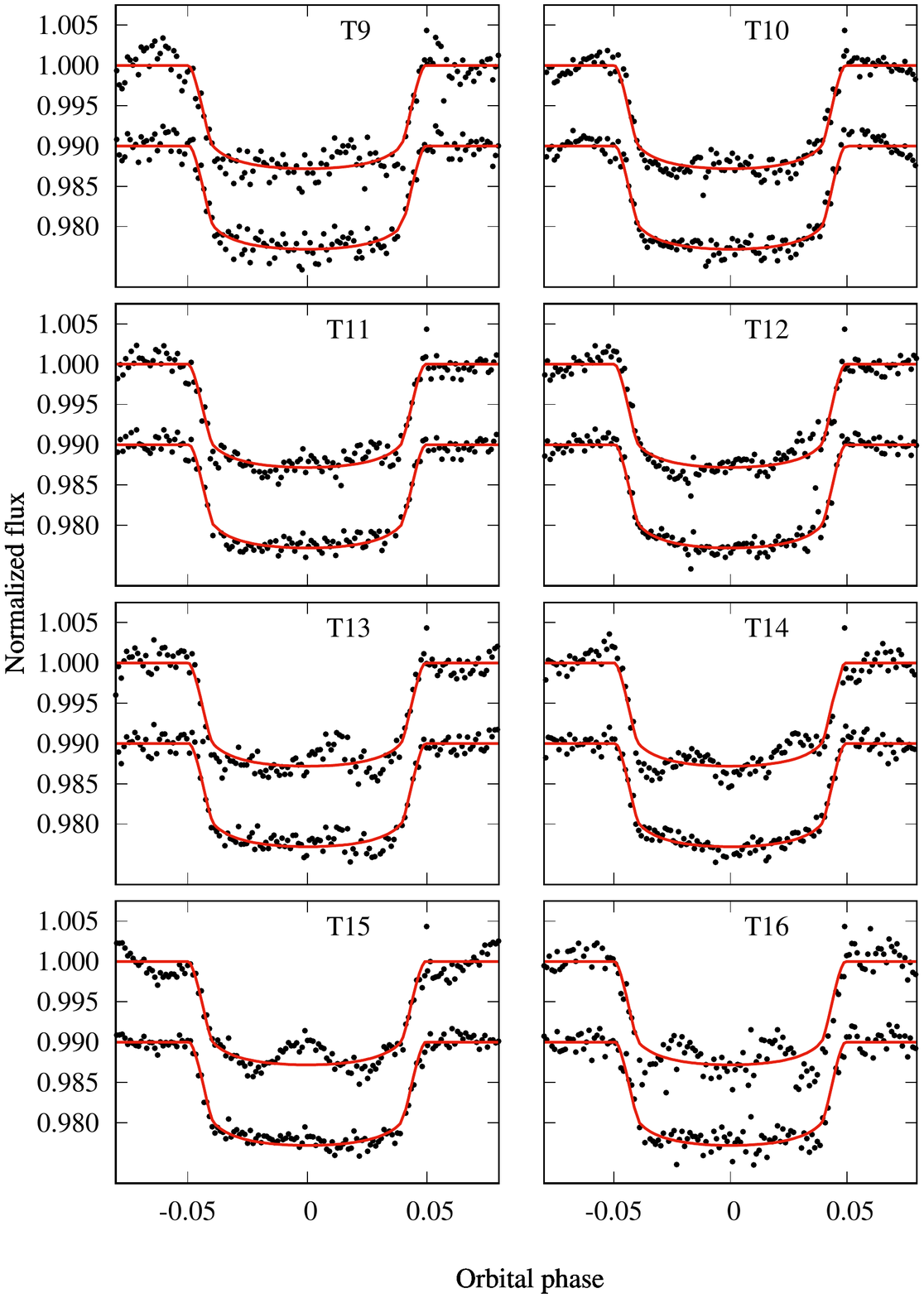}
    \caption{\label{fig:transits2} Same as Fig.~\ref{fig:transits1}, but for the remaining 8 primary transits.}
\end{figure*}

\section{Posterior distributions} 

\begin{figure*}[ht!]
    \centering
    \includegraphics[width=\textwidth]{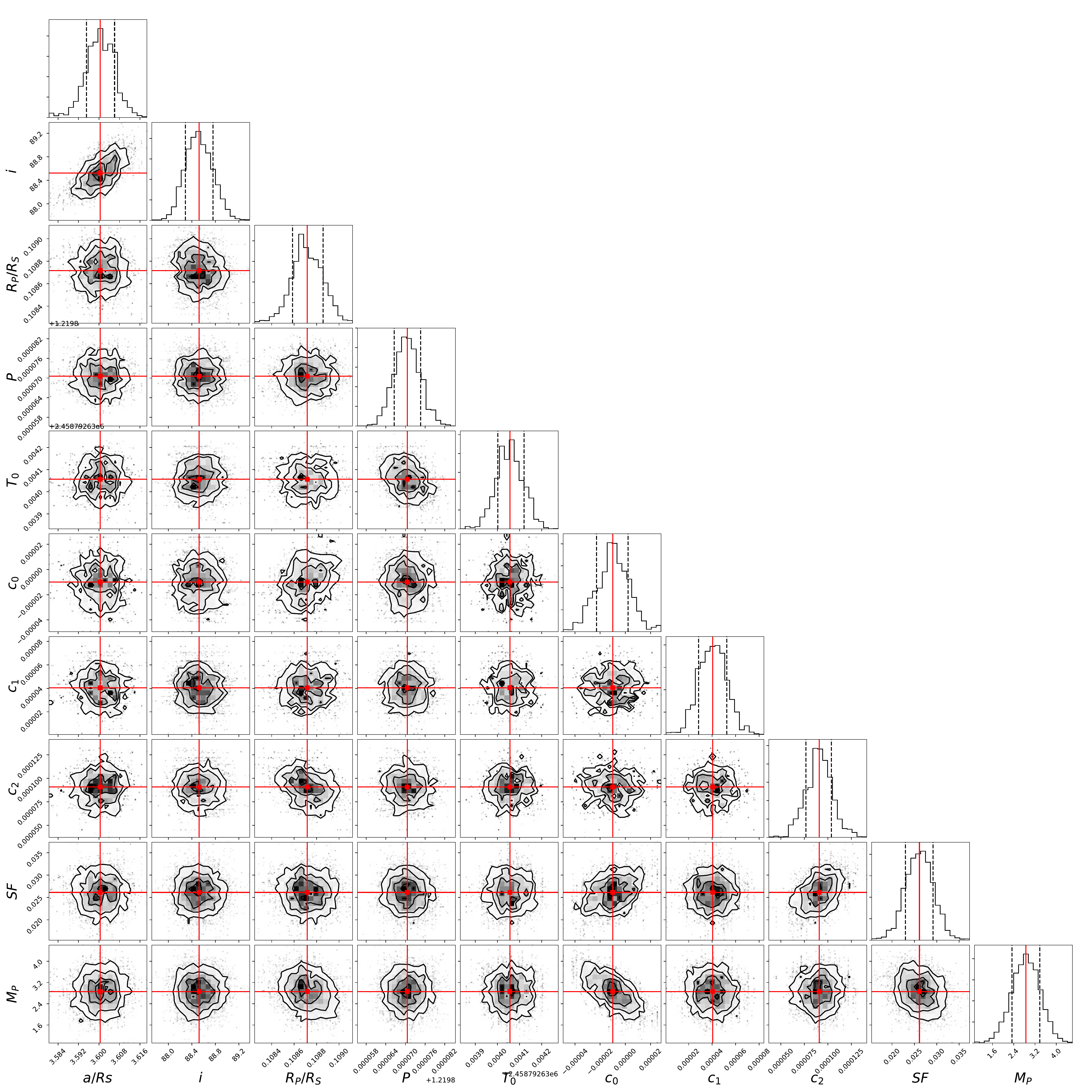}
    \caption{\label{fig:posteriors_MCMC} Posterior distributions for the fitted parameters specified in Table~\ref{tab:results_MCMC}, specifically for M4.}
\end{figure*}

\end{appendix}

\end{document}